\newcommand{\CLASSINPUTtoptextmargin}{0.85in}
\newcommand{\CLASSINPUTbottomtextmargin}{0.97in}
\documentclass[conference]{IEEEtran}
\IEEEoverridecommandlockouts
\usepackage{standalone}%

\usepackage{cite}
\usepackage{amsmath,amssymb,amsfonts}
\usepackage{mathtools}
\usepackage{bm}
\usepackage{algorithmic}
\usepackage{graphicx}
\usepackage{textcomp}
\usepackage{subcaption}
\usepackage{xcolor}
\usepackage{trfsigns}
\usepackage{placeins}
\graphicspath{{inkscape/}} %
\usepackage{tabularx}         
\usepackage{multirow}  
\usepackage{hhline}                    
\usepackage[shortcuts,acronym]{glossaries}
\usepackage{blkarray, bigstrut}

\def\BibTeX{{\rm B\kern-.05em{\sc i\kern-.025em b}\kern-.08em T\kern-.1667em\lower.7ex\hbox{E}\kern-.125emX}}

\usepackage{pgfplots}
\pgfplotsset{compat=newest} %
\usepackage{tikz}
\usetikzlibrary{calc}

\pgfkeys{
    /includestandaloneresized/.is family, /includestandaloneresized,
    default/.style = {
        width = \textwidth,
        ratio = 0.61803398875},
    width/.estore in = \itgWidth,
    ratio/.estore in = \itgRatio
}
\newlength\figurewidth
\newlength\figureheight

\def\unit{0.78cm}  %
\usetikzlibrary{shapes.geometric, positioning, calc, shapes.arrows, backgrounds}
\usepackage{amsmath,amssymb,amsfonts}
\usetikzlibrary{shapes, shapes.geometric, positioning, calc, shapes.arrows, backgrounds, shapes.gates.logic.US, shapes.gates.logic.IEC}

\usepgfplotslibrary{groupplots,dateplot}
\newenvironment{customlegend}[1][]{%
    \begingroup
    \csname pgfplots@init@cleared@structures\endcsname
    \pgfplotsset{#1}%
}{%
    \csname pgfplots@createlegend\endcsname
    \endgroup
}%
\def\addlegendimage{\csname pgfplots@addlegendimage\endcsname}

\tikzset{
    pics/twoarrows/.style ={code={
            \draw[-latex](+0.25,0.15)--(-0.25,0.15);
            \draw[-latex](-0.25,-0.15)--(+0.25,-0.15);
    }}
}

\makeatletter
\def\ps@IEEEtitlepagestyle{%
    \def\@oddfoot{\mycopyrightnotice}%
    \def\@evenfoot{}%
}
\def\mycopyrightnotice{%
    {\footnotesize
        This work has been submitted to the IEEE for possible publication. Copyright may be transferred without notice, after which this version may no longer be accessible.\hfill}%
    \gdef\mycopyrightnotice{}%
}

\IEEEoverridecommandlockouts\IEEEpubid{\makebox[\columnwidth]{\hfill} \hspace{\columnsep}\makebox[\columnwidth]{ }}
\begin{document}
\setlength\abovecaptionskip{0.45\baselineskip}
\setlength{\textfloatsep}{0.35\baselineskip}
\def\@IEEEfigurecaptionsepspace{\vskip\abovecaptionskip\relax}%
\def\@IEEEtablecaptionsepspace{\vskip\abovecaptionskip\relax}%
\title{A Variable Node Design with Check Node Aware Quantization Leveraging 2-Bit LDPC Decoding}

\author{
\IEEEauthorblockN{Philipp Mohr, Gerhard Bauch}
\IEEEauthorblockA{\textit{Hamburg University of Technology} \\
\textit{Institute of Communications}\\
21073 Hamburg, Germany \\
Email: \{philipp.mohr, bauch\}@tuhh.de}
}

\maketitle

\newacronym{app}{APP}{a-posteriori probability}
\newacronym{bp}{BP}{belief propagation}
\newacronym{llr}{LLR}{log-likelihood ratio}
\newacronym{lut}{LUT}{lookup table}
\newacronym{luts}{LUTs}{lookup tables}
\newacronym{ib}{IB}{information bottleneck}
\newacronym{ldpc}{LDPC}{low-density parity-check}
\newacronym{qc}{QC}{quasi-cyclic}
\newacronym{omsq}{OMSQ}{quantized offset-min-sum}
\newacronym[longplural={variable nodes}]{vn}{VN}{variable node}
\newacronym[longplural={check nodes}]{cn}{CN}{check node}

\begin{abstract}
For improving coarsely quantized decoding of LDPC codes, we propose a check node aware design of the variable node update. In contrast to previous works, we optimize the variable node to explicitly maximize the mutual information preserved in the check-to-variable instead of the variable-to-check node messages. The extended optimization leads to a significantly different solution for the compression operation at the variable node. Simulation results for regular LDPC codes confirm that the check node aware design, especially for very coarse quantization with 2- or 3-bit messages, achieves performance gains of up to 0.2 dB - without additional hardware costs. We also show that the 2-bit message resolution enables a very efficient implementation of the check node update, which requires only 2/9 of the 3-bit check node's transistor count and reduces the signal propagation delay by a factor of 4.
\end{abstract}

\newcommand\myvec{\boldsymbol}%
\newcommand\mymat[1]{\boldsymbol{\mathbf{#1}}}
\newcommand\myup{\mathrm}
\newcommand\mysamplespace{\mathcal}
\newcommand\mycard[1]{\lvert #1\rvert}
\newcommand\myset[1]{\left\{#1\right\}}
\newcommand\mynotset[1]{\sim\{#1\}}
\newcommand\mylabel{\mathrm}
\newcommand\myvar[1]{\mathsf{#1}}
\FloatBarrier
\section{Introduction}\label{sec:introduction}
Low-resolution message passing plays a key role in hardware efficient implementations of low-density parity-check (LDPC) decoders. In particular the transfer of message bits between variable and check nodes causes high energy and chip area consumption. Hence, the node updates within practical implementations include quantization operations. In recent years, finite alphabet decoders \cite{lewandowsky_trellis_2015, meidlinger_quantized_2015, he_mutual_2019,mohr_coarsely_2021, wang_reconstruction-computation-quantization_2022,mohr_uniform_2022} that maximize the mutual information in the decoding process, have shown superior performance over conventionally designed decoders like the normalized or offset min-sum algorithms\cite{jinghu_chen_reduced-complexity_2005}. The mutual information maximizing decoders are typically optimized in an offline design phase, where discrete density evolution tracks the joint distribution of messages and relevant variables\cite{kurkoski_noise_2008}.

The objective of maximizing the mutual information can be solved for different local optimization levels:
One option is a lookup table design that decomposes each node into concatenated two-input lookup tables\cite{lewandowsky_trellis_2015, meidlinger_quantized_2015}. To avoid increasing internal resolutions that would cause non-practical table sizes, each lookup table performs a compression operation that is optimized to maximize the preserved mutual information in the output message. 

Another option is a computational domain technique which improves the performance by allowing higher internal resolutions in the node update\cite{he_mutual_2019}. In that approach, all input messages are translated to higher resolution representation values (e.g. proportional to log-likelihood ratios) and merged with arithmetic operations. Finally, a quantization operation maximizes the mutual information in the output message.

Yet, the two aforementioned classes of discrete decoders neglect the check node behavior when optimizing the compression operation at the variable node. An important insight about the check node update is that the reliability of the output message is similar to the lowest reliability found in the extrinsic input messages. Hence, input messages that represent high reliability have a very low chance to affect the output message of the check node significantly. 
Ultimately, the decoder performance depends on the quality of the check node and channel messages which are processed in the variable node to compute the hard decisions. This implies potential for improvement by designing the variable node's compression operation with the overall goal to maximize the preserved mutual information \emph{after} the check node update.
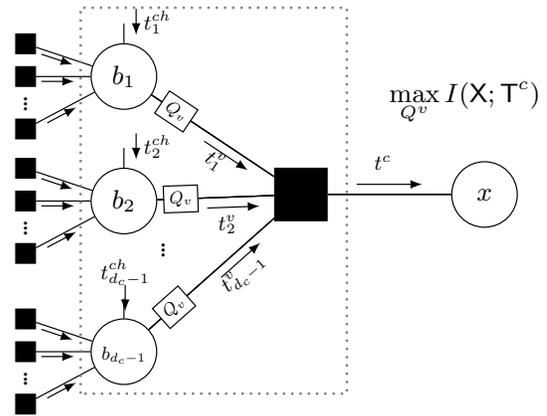
\begin{figure}[t]
	\centering
    \small
\begin{tikzpicture} [scale=1.0, x=\unit,y=\unit, varnode/.style={circle, inner sep=0pt, },  node distance=1.5*\unit and 1.0*\unit]

\node(cn0)[rectangle, minimum width=0.8*\unit,minimum height=0.8*\unit,draw=black, fill=black]{};
\node(vn0)[varnode, minimum height=1.0*\unit, draw=black, right of=cn0, xshift=1.3*\unit]{$x$};
\draw(vn0)--(cn0);

\node(vn1)[varnode, minimum height=1.0*\unit, draw=black, left of=cn0, yshift=1.8*\unit, xshift=-1.2*\unit]{$b_1$};
\node(cn11)[rectangle, minimum width=0.3*\unit,minimum height=0.3*\unit,draw=black, fill=black, left of=vn1, yshift=0.5*\unit]{};
\node(cn12)[rectangle, minimum width=0.3*\unit,minimum height=0.3*\unit,draw=black, fill=black, below of=cn11, yshift=1.0*\unit]{};
\node(cn13)[rectangle, minimum width=0.3*\unit,minimum height=0.3*\unit,draw=black, fill=black, below of=cn12,  yshift=0.7*\unit]{};
\node (dotscns1) at ($(cn12)!0.45!(cn13)$)[scale=0.4]{\Huge${\vdots}$};
\draw($(vn1.north)+(0,0.3*\unit)$)--node[right, sloped, xshift=-0.6*\unit, yshift=0.5*\unit]{\tikz\draw[-latex](-0.2,0)--node[right, rotate=90,scale=0.8]{$t^{ch}_{1}$}(0.2,0);}(vn1.north);
\draw(cn11)--node[below, sloped, pos=0.4, yshift=0.08*\unit]{\tikz\draw[-latex](-0.25,0)--node[below,scale=0.8]{}(0.25,0);}(vn1);
\draw(cn12)--node[below, sloped, pos=0.4, yshift=0.08*\unit]{\tikz\draw[-latex](-0.25,0)--node[below,scale=0.8]{}(0.25,0);}(vn1);
\draw(cn13)--node[below, sloped, pos=0.4, yshift=0.08*\unit]{\tikz\draw[-latex](-0.25,0)--node[below,scale=0.8]{}(0.25,0);}(vn1);
\draw(vn1)--(cn0);

\node(vn2)[varnode, minimum height=1.0*\unit, draw=black, below of=vn1, yshift=-0.4*\unit]{$b_2$};
\node(cn21)[rectangle, minimum width=0.3*\unit,minimum height=0.3*\unit,draw=black, fill=black, left of=vn2, yshift=0.5*\unit]{};
\node(cn22)[rectangle, minimum width=0.3*\unit,minimum height=0.3*\unit,draw=black, fill=black, below of=cn21, yshift=1.0*\unit]{};
\node(cn23)[rectangle, minimum width=0.3*\unit,minimum height=0.3*\unit,draw=black, fill=black, below of=cn22,  yshift=0.7*\unit]{};
\node (dotscns2) at ($(cn22)!0.45!(cn23)$)[scale=0.4]{\Huge${\vdots}$};
\draw($(vn2.north)+(0,0.3*\unit)$)--node[right, sloped, xshift=-0.6*\unit, yshift=0.5*\unit]{\tikz\draw[-latex](-0.2,0)--node[right, rotate=90,scale=0.8]{$t^{ch}_{2}$}(0.2,0);}(vn2.north);
\draw(cn21)--node[below, sloped, pos=0.4, yshift=0.08*\unit]{\tikz\draw[-latex](-0.25,0)--node[below,scale=0.8]{}(0.25,0);}(vn2);
\draw(cn22)--node[below, sloped, pos=0.4, yshift=0.08*\unit]{\tikz\draw[-latex](-0.25,0)--node[below,scale=0.8]{}(0.25,0);}(vn2);
\draw(cn23)--node[below, sloped, pos=0.4, yshift=0.08*\unit]{\tikz\draw[-latex](-0.25,0)--node[below,scale=0.8]{}(0.25,0);}(vn2);
\draw(vn2)--(cn0);

\node(vn3)[varnode, minimum height=1.0*\unit, draw=black, below of=vn2, yshift=-0.8*\unit, label={[yshift=-0.8*\unit,scale=0.7]$b_{d_c{-}1}$}]{};
\node(cn31)[rectangle, minimum width=0.3*\unit,minimum height=0.3*\unit,draw=black, fill=black, left of=vn3, yshift=0.5*\unit]{};
\node(cn32)[rectangle, minimum width=0.3*\unit,minimum height=0.3*\unit,draw=black, fill=black, below of=cn31, yshift=1.0*\unit]{};
\node(cn33)[rectangle, minimum width=0.3*\unit,minimum height=0.3*\unit,draw=black, fill=black, below of=cn32,  yshift=0.7*\unit]{};
\node (dotscns3) at ($(cn32)!0.45!(cn33)$)[scale=0.4]{\Huge${\vdots}$};
\draw($(vn3.north)+(0,0.3*\unit)$)--node[below, sloped, xshift=-0.4*\unit, yshift=0.65*\unit]{\tikz\draw[-latex](-0.2,0)--node[above, rotate=90,scale=0.8, yshift=0.1*\unit]{$t^{ch}_{d_c-1}$}(0.2,0);}(vn3.north);
\draw(cn31)--node[below, sloped, pos=0.4, yshift=0.08*\unit]{\tikz\draw[-latex](-0.25,0)--node[below,scale=0.8]{}(0.25,0);}(vn3);
\draw(cn32)--node[below, sloped, pos=0.4, yshift=0.08*\unit]{\tikz\draw[-latex](-0.25,0)--node[below,scale=0.8]{}(0.25,0);}(vn3);
\draw(cn33)--node[below, sloped, pos=0.4, yshift=0.08*\unit]{\tikz\draw[-latex](-0.25,0)--node[below,scale=0.8]{}(0.25,0);}(vn3);
\draw(vn3)--(cn0);

\node (dotsvns) at ($(vn2)!0.3!(vn3)$)[scale=0.4, xshift=1.5*\unit]{\Huge${\vdots}$};

\draw (vn1)--node[below, sloped, pos=0.65, yshift=0.1*\unit]{\tikz\draw[-latex](-0.4,0)--node[below, rotate=0, xshift=-0.1*\unit,yshift=0.1*\unit,scale=0.8]{$t^{v}_{1}$}(0.4,0);}(cn0);
\draw (vn1)--node[sloped, pos=0.2]{\tikz\node[rectangle, draw=black, fill=white, scale=0.65]{$Q_v$};}(cn0);

\draw (vn2)--node[below, sloped, pos=0.65, yshift=0.0*\unit]{\tikz\draw[-latex](-0.4,0)--node[below, rotate=0, xshift=-0.1*\unit,yshift=0.0*\unit,scale=0.8]{$t^{v}_{2}$}(0.4,0);}(cn0);
\draw (vn2)--node[sloped, pos=0.2]{\tikz\node[rectangle, draw=black, fill=white, scale=0.65]{$Q_v$};}(cn0);

\draw (vn3)--node[below, sloped, pos=0.65, yshift=0.1*\unit]{\tikz\draw[-latex](-0.4,0)--node[below, rotate=0, xshift=-0.1*\unit,yshift=0.1*\unit,scale=0.8]{$t^{v}_{d_c-1}$}(0.4,0);}(cn0);
\draw (vn3)--node[sloped, pos=0.2]{\tikz\node[rectangle, draw=black, fill=white, scale=0.65]{$Q_v$};}(cn0);

\draw[gray,thick,dotted] ($(vn3.north west)+(-0.3*\unit,-1.0*\unit)$)  rectangle ($(cn0|-vn1)+(0.7*\unit,1.05*\unit)$);
\node(maxmi) at ($(vn0)+(-0.3*\unit,1.4*\unit)$){$\begin{aligned}
    \max_{Q^v} I(\mathsf{X};\mathsf{T}^c)
    \end{aligned}$};
\draw (cn0)--node[midway, above, sloped]{\tikz\draw[-latex](-0.5,0)--node[above, rotate=0, xshift=-0.1*\unit,yshift=0.1*\unit,scale=0.8]{$t^{c}$}(0.5,0);}(vn0);

\end{tikzpicture}
	\caption{A check node aware design of the variable node's compression operation. 
    }
	\label{fig:level3optimization}
\end{figure}

In this paper, we further extend the local optimization scope to a third level where the variable and check nodes (circle and squares, respectively) are jointly considered in the optimization, as depicted in Fig.~\ref{fig:level3optimization}. We showed in\cite{mohr_uniform_2022} that the restriction to uniformly placed thresholds achieved similar performance as non-uniform quantization at significantly reduced hardware costs. In addition, the check node with the widely applied minimum approximation is an excellent performance-complexity trade-off\cite{mohr_uniform_2022,meidlinger_quantized_2015}. Therefore, a combination of uniform quantization in the variable node and minimum approximation in the check node is considered here as a very appealing choice for hardware implementations. Moreover, we exploit that this configuration requires a single-parameter optimization: Only the spacing between the uniform boundaries (alternatively maximum representation range) must be adjusted, which reduces the computational effort of the joint optimization of variable and check nodes. Further, we note that under symmetric 2-bit decoding, the solutions with uniform and non-uniform quantization are equivalent.
The contributions of this paper can be summarized as follows.
\begin{itemize}
    \item A check node aware design of the variable node, that explicitly maximizes the mutual information in the check node messages, is performed for the first time, to the best of our knowledge.
    \item We analyze the performance in terms of mutual information and error rates in detail. Especially decoding under very coarse quantization benefits from the proposed method. We reveal significant gains of up to 0.2\,dB for 2-bit decoding. 
    \item We show that a 2-bit check node update requires only 2/9 of the logic gate count and 1/4 of the propagation delay compared to a 3-bit check node update.
\end{itemize}

Next, section II presents the system structure. In particular, check and variable node operations are described in detail. Furthermore, the hardware complexity of 2-bit and 3-bit symmetric check node processing is compared. Section III explains the check node aware design of the variable node and evaluates the performance in terms of mutual information. Finally, section IV compares the error rate performance for different regular LDPC codes.%

\section{System Structure}%
We assume a binary LDPC code with parity check matrix $\mymat{H}\in \{0,1\}^{N_c\times N}$ which can be represented by a Tanner graph with $N$ \glspl{vn} and $N_c$ \glspl{cn}. The encoder maps the information bits $\myvec{u}=[u_1, \ldots,u_{K}]$ to code bits $\myvec{b}=[b_1, \ldots, b_{N}]$ satisfying $\mymat{H}\myvec{b}^T=\myvec{0}$. For the transmission, we consider binary phase-shift keying (BPSK) symbols which are disturbed by additive white Gaussian noise (AWGN) at the receiver. A mutual information maximizing symmetric channel quantizer (like in \cite{lewandowsky_trellis_2015}) maps the received symbols to \mbox{$w_{ch}$-bit} messages $\myvec{t}^{ch}{\in}\mysamplespace{T}_{w_{ch}}^N$ with finite sign-magnitude alphabet $\mysamplespace{T}_l=\{-2^{l-1},\ldots,-1,{+}1,\ldots,{+}2^{l-1}\}$. In the decoder, $w$-bit messages are exchanged over multiple decoding iterations between variable and check nodes.
We apply a flooding schedule to perform the node updates. However, the proposed techniques can also be adopted for other schedules (e.g. layered scheduling\cite{mohr_coarsely_2021}). In the first decoding iteration the channel messages are directly mapped to the variable node output. If $w{<}w_{ch}$ an initial variable node update is designed to perform this compression mapping, which can be implemented by a small single-input lookup table.
\subsection{Check Node Update with Minimum Approximation}
A check node with degree $d_c$ exploits its underlying parity check equation $b_1\oplus b_2\oplus\ldots\oplus b_{d_c}=0$ to compute extrinsic information about the participating code bits. The incoming \mbox{$w$-bit} variable node messages $t_i^v\in \mysamplespace{T}_w, i=1,\ldots,d_c$ encode the probabilities $p(b_i|t^v_i)$ which can also be  represented by log-probability ratios $L_{t^v_i}=\log\frac{p(b_i=0|t^v_i)}{p(b_i=1|t^v_i)}$. 

\begin{figure}[t]
    \centering
    \small
\begin{tikzpicture} [scale=1.0, x=\unit,y=\unit, varnode/.style={circle, inner sep=0pt, minimum height=1.0*\unit, draw=black},  checknode/.style={rectangle, inner sep=0pt, minimum width=0.8*\unit,minimum height=0.8*\unit,draw=black, fill=black}, node distance=1.5*\unit and 1.0*\unit]

\node(vnx3)[varnode, minimum height=1.0*\unit, draw=black, xshift=1.3*\unit]{$x_3$};
\node(cn11)[checknode, above of=vnx3, xshift=-3*\unit]{};
\node(cn12)[checknode, above of=vnx3, xshift=3*\unit]{};

\node(vnx1)[varnode, above of=cn11, xshift=-1.4*\unit]{$x_1$};
\node(vnx2)[varnode, above of=cn11, xshift=1.4*\unit]{$x_2$};
\node(vnx4)[varnode, above of=cn12, xshift=-1.4*\unit]{$x_4$};
\node(vnx5)[varnode, above of=cn12, xshift=1.4*\unit]{$x_5$};

\node(cn01)[checknode, above of=vnx1]{};
\node(cn02)[checknode, above of=vnx2]{};
\node(cn03)[checknode, above of=vnx4]{};
\node(cn04)[checknode, above of=vnx5]{};

\node(vn0)[varnode, above of=cn01, xshift=-0.7*\unit, yshift=0.5*\unit]{$b_1$};
\node(vn1)[varnode, above of=cn01, xshift=0.7*\unit, yshift=0.5*\unit]{$b_2$};
\node(vn2)[varnode, above of=cn02, xshift=-0.7*\unit, yshift=0.5*\unit]{$b_3$};
\node(vn3)[varnode, above of=cn02, xshift=0.7*\unit, yshift=0.5*\unit]{$b_4$};

\node(vn4)[varnode, above of=cn03, xshift=-0.7*\unit, yshift=0.5*\unit]{$b_5$};
\node(vn5)[varnode, above of=cn03, xshift=0.7*\unit, yshift=0.5*\unit]{$b_6$};
\node(vn6)[varnode, above of=cn04, xshift=-0.7*\unit, yshift=0.5*\unit]{$b_7$};
\node(vn7)[varnode, above of=cn04, xshift=0.7*\unit, yshift=0.5*\unit]{$b_8$};

\draw(vnx3)--pic[sloped, pos=0.5]{twoarrows}(cn11.south);
\draw(cn11.north)--pic[sloped, pos=0.5]{twoarrows}(vnx1);
\draw(cn11.north)--pic[sloped, pos=0.5]{twoarrows}(vnx2);

\draw(vnx3)--pic[sloped, pos=0.5]{twoarrows}(cn12.south);
\draw(cn12.north)--pic[sloped, pos=0.5]{twoarrows}(vnx4);
\draw(cn12.north)--pic[sloped, pos=0.5]{twoarrows}(vnx5);

\draw(vnx1)--node[sloped, pos=0.5]{\begin{tikzpicture}
    \draw[-latex,red](+0.25,0.15)--(-0.25,0.15);
    \draw[-latex](-0.25,-0.15)--(+0.25,-0.15);
    \end{tikzpicture}}(cn01.south);
\draw(cn01.north)--node[sloped, pos=0.6]{\begin{tikzpicture}
    \draw[-latex](+0.25,0.15)--(-0.25,0.15);
    \draw[-latex,red](-0.25,-0.15)--(+0.25,-0.15);
    \end{tikzpicture}}(vn0.south);
\draw(cn01.north)--node[sloped, pos=0.6]{\begin{tikzpicture}
    \draw[-latex, red](+0.25,0.15)--(-0.25,0.15);
    \draw[-latex](-0.25,-0.15)--(+0.25,-0.15);
    \end{tikzpicture}}(vn1.south);
\node(labelyv0)[below of=vn0, red, xshift=-0.15*\unit, yshift=0.6*\unit]{$t_1^v$};
\node(labelyv1)[below of=vn1, red, xshift=-0.69*\unit, yshift=0.8*\unit]{$t_2^v$};
\node(labelt0)[below of=cn01, red, xshift=-0.5*\unit, yshift=0.7*\unit]{$t$};

\draw(vnx2)--pic[sloped, pos=0.5]{twoarrows}(cn02.south);
\draw(cn02.north)--pic[sloped, pos=0.6]{twoarrows}(vn2.south);
\draw(cn02.north)--pic[sloped, pos=0.6]{twoarrows}(vn3.south);

\draw(vnx4)--pic[sloped, pos=0.5]{twoarrows}(cn03.south);
\draw(cn03.north)--pic[sloped, pos=0.6]{twoarrows}(vn4.south);
\draw(cn03.north)--pic[sloped, pos=0.6]{twoarrows}(vn5.south);

\draw(vnx5)--pic[sloped, pos=0.5]{twoarrows}(cn04.south);
\draw(cn04.north)--pic[sloped, pos=0.6]{twoarrows}(vn6.south);
\draw(cn04.north)--pic[sloped, pos=0.6]{twoarrows}(vn7.south);

\draw[black,thick] ($(cn01.north west)-(0.3*\unit,-0.3*\unit)$)  rectangle ($(cn04|-vnx3)-(-0.7*\unit,0.6*\unit)$);

\end{tikzpicture}
    \caption{Full check node update ($d_c{=}8$) in a tree structure.}
    \vspace{-0.5cm}
    \label{fig:checknodetree}
\end{figure}
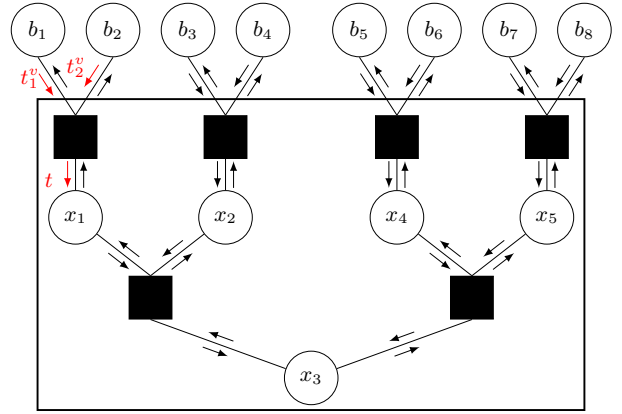

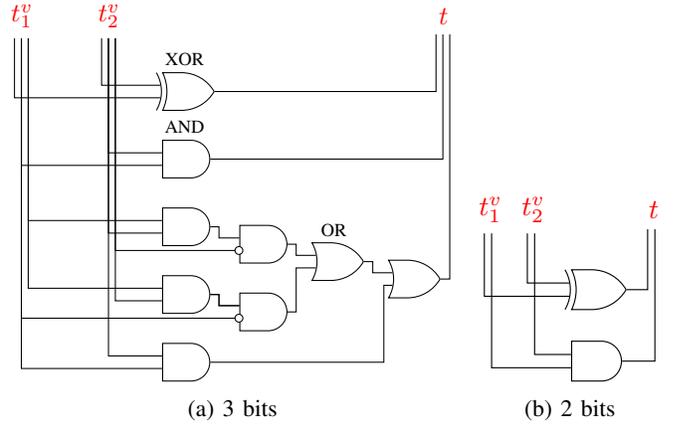
\begin{figure}[t]
    \begin{subfigure}[b]{0.69\linewidth}
            \begin{tikzpicture} [scale=1.0, x=\unit,y=\unit, node distance=0.5*\unit and 0.5*\unit,
    invert/.style={circle, left, xshift=0.07*\unit, minimum size=3pt, inner sep=0pt,draw=black, fill=white}]      
    \node(c)[red]{$t$};
    \node(c0)[below=of c, yshift=0.8*\unit, xshift=-0.13*\unit]{};
    \node(c1)[right=of c0, xshift=-0.7*\unit]{};
    \node(c2)[right=of c1, xshift=-0.7*\unit]{};
    
    \node(xor0)[xor gate US, draw, logic gate inputs=nn, left=of c, xshift=-3.2*\unit, yshift=-1.3*\unit]{};
    \node(and1)[and gate US, draw, logic gate inputs=nn, below=of xor0]{};
    
    \node(and200)[and gate US, draw, logic gate inputs=nn, below=of and1]{};
    \node(and210)[and gate US, draw, logic gate inputs=nn, below=of and200]{};
    \node(and220)[and gate US, draw, logic gate inputs=nn, below=of and210]{};
    
    \node(and201)[and gate US, draw, logic gate inputs=nn, right=of and200, yshift=-0.3*\unit]{};
    \node(and211)[and gate US, draw, logic gate inputs=nn, below=of and201]{};
    
    \node(or20)[or gate US, draw, logic gate inputs=nn, right=of and201, yshift=-0.3*\unit]{};
    \node(or21)[or gate US, draw, logic gate inputs=nn, right=of or20, yshift=-0.3*\unit]{};
    
    \node(a)[red, left of=xor0,xshift=-2.3*\unit, yshift=1.3*\unit]{$t^v_1$};
    \node(a0)[below=of a, yshift=0.8*\unit, xshift=-0.13*\unit]{};
    \node(a1)[right=of a0, xshift=-0.7*\unit]{};
    \node(a2)[right=of a1, xshift=-0.7*\unit]{};
    
    \node(b)[red,right of=a, xshift=1*\unit]{$t^v_2$};
    \node(b0)[below=of b, yshift=0.8*\unit, xshift=-0.13*\unit]{};
    \node(b1)[right=of b0, xshift=-0.7*\unit]{};
    \node(b2)[right=of b1, xshift=-0.7*\unit]{};
    
    \draw(a0)|-(xor0.input 2);
    \draw(b0)|-(xor0.input 1);
    \draw(xor0.output)-|(c0);
    
    \draw(a1)|-(and1.input 2);
    \draw(b1)|-(and1.input 1);
    \draw(and1.output)-|(c1);
    
    \draw(a2)|-(and200.input 1);
    \draw(b1)|-(and200.input 2); \draw(and200.output)--+(0.15,0)|-(and201.input 1);
    \draw(b2)|-(and201.input 2);
    \node at (and201.input 2) [invert]{};
    \draw(a2)|-(and210.input 1);
    \draw(b2)|-(and210.input 2); \draw(and210.output)--+(0.15,0)|-(and211.input 1);
    \draw(a1)|-(and211.input 2);
    \node at (and211.input 2) [invert]{}; 
    \draw(b1)|-(and220.input 1);
    \draw(a1)|-(and220.input 2); 
    \draw(and210.output)--+(0.15,0)|-(and211.input 1);
    \draw(and201.output)--+(0.15,0)|-(or20.input 1);
    \draw(and211.output)--+(0.15,0)|-(or20.input 2);
    \draw(or20.output)--+(0.15,0)|-(or21.input 1);
    \draw(or21.input 2)--+(-0.15,0)|-(and220.output);
    
    \draw(or21.output)-|(c2);

    \node(labelxor)[above= of xor0, yshift=-0.5*\unit, scale=0.7]{XOR};
    \node(labeland)[above= of and1, yshift=-0.5*\unit, scale=0.7]{AND};
    \node(labelor)[above= of or20, yshift=-0.5*\unit, scale=0.7]{OR};
    \end{tikzpicture}
        \caption{3 bits}
        \label{fig:checknode3bit}
    \end{subfigure}
    \hfill
    \begin{subfigure}[b]{0.3\linewidth}
            \begin{tikzpicture} [scale=1.0, x=\unit,y=\unit, node distance=0.5*\unit and 0.5*\unit,
    invert/.style={circle, left, xshift=0.07*\unit, minimum size=3pt, inner sep=0pt,draw=black, fill=white}]      
    \node(t)[red]{$t$};
    \node(tb0)[below=of t, yshift=0.8*\unit, xshift=-0.1*\unit]{};
    \node(tb1)[right=of tb0, xshift=-0.7*\unit]{};
    
    \node(xor0)[xor gate US, draw, logic gate inputs=nn, left=of t, xshift=0.3*\unit, yshift=-1.3*\unit]{};
    \node(and1)[and gate US, draw, logic gate inputs=nn, below=of xor0]{};
    
    \node(y0)[red, left of=xor0,xshift=-1.2*\unit, yshift=1.3*\unit]{$t^v_1$};
    \node(y0b0)[below=of y0, yshift=0.8*\unit, xshift=-0.1*\unit]{};
    \node(y0b1)[right=of y0b0, xshift=-0.7*\unit]{};
    
    \node(y1)[red,right of=y0, xshift=0.2*\unit]{$t^v_2$};
    \node(y1b0)[below=of y1, yshift=0.8*\unit, xshift=-0.1*\unit]{};
    \node(y1b1)[right=of y1b0, xshift=-0.7*\unit]{};
    
    \draw(y0b0)|-(xor0.input 2);
    \draw(y1b0)|-(xor0.input 1);
    \draw(xor0.output)-|(tb0);
    
    \draw(y0b1)|-(and1.input 2);
    \draw(y1b1)|-(and1.input 1);
    \draw(and1.output)-|(tb1);

    \end{tikzpicture}
        \caption{2 bits}
        \label{fig:checknode2bit}
    \end{subfigure}
    \caption{Circuits for a two-input check node update.}
    \label{fig:checknodes32bit}
\end{figure}
A check node with $d_c>3$ inputs can be decomposed into multiple degree-three check nodes\cite{kschischang_factor_2001}. Minimum delay for the full check node update is achieved with a tree structure like it is shown in Fig.~\ref{fig:checknodetree}. With proper scheduling, each factor node computes only three output messages. E.g., extrinsic probability information for the auxiliary variable $x_1$ is provided by
\begin{align}
    p(x_1, t^v_1, t^v_2)=\sum_{b_1,b_2}p(x_1|b_2,b_1)p(b_1|t^v_1)p(b_2|t^v_2)
    \label{equ:cn_prob}
\end{align}
where $p(x_1|b_2,b_1)$ is obtained from the parity check equation $x_1=b_1\oplus b_2$. Performing the computation (\ref{equ:cn_prob}) in log-domain with $L\coloneqq \log\frac{p(x_1{=}0,t^v_1,t^v_2)}{p(x_1{=}1,t^v_1,t^v_2)}$ results in the well-known box-plus operation\cite{xiao-yu_hu_efficient_2001}
\begin{align}
\begin{split}
    L=L_{t^v_1}\boxplus L_{t^v_2}=&\operatorname{sgn}(L_{t^v_1})\operatorname{sgn}(L_{t^v_2})\min(|L_{t^v_1}|, |L_{t^v_2}|)\\
    &+f(|L_{t^v_1}|,|L_{t^v_2}|),
    \label{equ:boxplus}
\end{split}
\end{align}
where $f(x,y)=\log(1+e^{-|x-y|})-\log(1+e^{-|x+y|})$ is a correction term. %
When neglecting the correction term and if the messages use a symmetric sign-magnitude format, (\ref{equ:boxplus}) can be simplified resulting in the minimum approximation\cite{meidlinger_quantized_2015}:
\begin{align}
\begin{split}
t=\operatorname{sgn}(t^v_1)\operatorname{sgn}(t^v_2)\min(|t^v_1|, |t^v_2|).
\label{equ:cn_min_approx}
\end{split}
\end{align}
In Fig.~\ref{fig:checknodes32bit} we depict two optimized hardware implementations of (\ref{equ:cn_min_approx}) for 2- and 3-bit alphabet sizes. The circuits implement the Boolean expressions of the minimized disjunctive normal form obtained with the Karnaugh map technique. 
The 3-bit check node requires 9 logic gates to perform the update. The critical path consists of 4 serial logic gate operations. On the other hand, the 2-bit check node requires only 2 logic gates which can operate in parallel. The minimum delay is reduced by a factor of 4. The price to be paid is a performance loss, however, the check node aware design proposed by the paper can reduce this loss significantly.

The overall number of comparisons in a tree structure is $3(d_c-2)$ assuming $d_c$ is a power of two. We note that the check node update could also be performed with a first and second minima search which requires only $d_c+\lceil\log{d_c}\rceil-2$ comparisons\cite{lee_low-complexity_2015}. Yet, our approach avoids usage of multiplexers and leads to less delay.

\FloatBarrier
\subsection{Variable Node Update with Uniform Quantization}
A variable node with degree $d_v$ deals with the messages $t^{ch}\in \mysamplespace{T}_{w_{ch}}$ and $t^{c}_k\in \mysamplespace{T}_w,k\in\{1,...,d_v\}$ from the channel and the $d_v$ connected check nodes, respectively. The messages are assumed to provide independent information about the underlying code bit $b$. 
For finite alphabet decoders, multiple implementation options exist for the variable node. Here, we decided for the computational domain structure\cite{he_mutual_2019} which potentially offers better performance than the lookup table decomposition technique\cite{lewandowsky_trellis_2015}. 
For example, extrinsic information for the first connected check node in terms of an LLR yields
\begin{align}
    L(b|t^{ch},t^{c}_2,\ldots,t^{c}_{d_v})=L(b)+L(t^{ch}|b)+\sum_{k=2}^{d_v}L(t^{c}_k|b)
\end{align}
where $L(t|b){=}\log\frac{p(t|b{=}0)}{p(t|b{=}1)}$. For finite alphabet decoders, the LLRs can be stored in translation tables $\phi^{ch}_i(t^{ch}){=}L(t^{ch}|b)$ and $\phi^c_i(t^c){=}L(t^c|b)$. Finally, the resulting LLR $L(b|t^{ch},t^{c}_2,\ldots,t^{c}_{d_v})$ must be quantized  to a $w$-bit message
\begin{align}
t^v_1 = Q^v\left(\phi^{ch}(t^{ch}) + \sum_{k=2}^{d_v} \phi^c_k(t^c_k)\right).
\label{equ:vn_update}
\end{align}
In Fig.~\ref{fig:full_vnu_hardware} a hardware implementation of the variable node update for all connected check nodes is depicted. We point out, that symmetric boundaries and translation tables are enforced to reduce computational and space complexity\cite{mohr_uniform_2022}. In practice, a translation table is used that outputs scaled LLRs represented by integer numbers $\phi_{\Delta}(t){=}\operatorname{sgn}(\phi(t)) \operatorname{min}(\lfloor\frac{1}{\Delta}|\phi(t)|{+}\frac{1}{2}\rfloor, 2^{w_{\phi}{-}1}{-}1)$ where $\phi_{\Delta}(t)\approx\frac{1}{\Delta} \phi(t)$. The step size $\Delta\in \mathbb{R}^+$ defines the internal resolution of the LLRs %
and $w_{\phi}$ is the translation bit width.
In this paper we restrict to uniform quantization with $r{\in}\mathbb{N}_0$:
\begin{align}
Q^v(y)=\operatorname{sgn}(y)\min\left(\lfloor |y|/2^r\rfloor+1,2^{w-1}\right)
\label{equ:uniform_quantization}
\end{align}
It was observed in \cite{mohr_uniform_2022} that uniform quantization reduces the preserved mutual information $I(\myvar{B};\myvar{T}^v_1)$ only insignificantly compared to the optimal non-uniform quantization\cite{kurkoski_quantization_2014,he_mutual_2019}. At the same time, a much simpler hardware structure, depicted in Fig.~\ref{fig:hardware_quant_uniform}, is possible: For an $r$-bit right shift and a subsequent clipping operation only a few logic gates are required. 

\begin{figure}
	\centering
	    \begin{tikzpicture} [scale=1.0, x=\unit,y=1.0*\unit, node distance=0.3*\unit and 1.0*\unit]   
    \node(sum)[rectangle, draw=black, fill=white, minimum width = 8*\unit]{$+$};
    
    \node(twoch)[rectangle, draw=black, above=of sum.175, yshift=0*\unit]{\scriptsize 2's};
    \node(two1)[rectangle, draw=black, above=of sum.170, yshift=0*\unit]{\scriptsize 2's};
    \node(two2)[rectangle, draw=black, above=of sum.70, yshift=0*\unit]{\scriptsize 2's};
    \node(twodv)[rectangle, draw=black, above=of sum.5, yshift=0*\unit]{\scriptsize 2's};
    
    \node(absphimch)[rectangle, draw=black, above=of twoch, yshift=0.0*\unit]{\footnotesize$|\phi_\Delta^{ch}|$};
    \node(absphim1)[rectangle, draw=black, above=of two1, yshift=0.0*\unit]{\footnotesize$|\phi_\Delta^c|$};
    \node(absphim2)[rectangle, draw=black, above=of two2, yshift=0.0*\unit]{\footnotesize$|\phi_\Delta^c|$};
    \node(absphimdv)[rectangle, draw=black, above=of twodv, yshift=0.0*\unit]{\footnotesize$|\phi_\Delta^c|$};
    
    \node(mch)[above=of absphimch, yshift=0*\unit]{$t^{ch}$};
    \node(m1)[above=of absphim1, yshift=0*\unit]{$t^c_1$};
    \node(m2)[above=of absphim2, yshift=0*\unit]{$t^c_{2}$};
    \node(mdv)[above=of absphimdv, yshift=0*\unit]{$t^c_{d_v}$};
    \node (mdots) at ($(m2)!0.5!(mdv)$) {$\dots$};
    
    \node(minus1)[rectangle, draw=black, below=of two1, yshift=-1*\unit]{$+$};
    \node(minus2)[rectangle, draw=black, below=of two2, yshift=-1*\unit]{$+$};
    \node(minusdv)[rectangle, draw=black, below=of twodv, yshift=-1*\unit]{$+$};
    
    \node(sm1)[rectangle, draw=black, below=of minus1, yshift=0*\unit]{\scriptsize SM};
    \node(sm2)[rectangle, draw=black, below=of minus2, yshift=0*\unit]{\scriptsize SM};
    \node(smdv)[rectangle, draw=black, below=of minusdv, yshift=0*\unit]{\scriptsize SM};
    
    \node(quant1)[rectangle, draw=black, below=of sm1, yshift=-0.2*\unit]{$Q$};
    \node(quant2)[rectangle, draw=black, below=of sm2, yshift=-0.2*\unit]{$Q$};
    \node(quantdv)[rectangle, draw=black, below=of smdv, yshift=-0.2*\unit]{$Q$};
    
    \node(mv1)[below=of quant1, yshift=-0.5*\unit]{$t^v_1$};
    \node(mv2)[below=of quant2, yshift=-0.5*\unit]{$t^v_2$};
    \node(mvdv)[below=of quantdv, yshift=-0.5*\unit]{$t^v_{d_v}$};
    \node (mvdots) at ($(mv2)!0.5!(mvdv)$) {$\dots$};
    
    \draw[-latex](mch.south)--(absphimch);
    \draw[-latex](absphimch.south)--(absphimch|-twoch.north);
    \draw[-latex](mch.south)|-+(0.75,-0.05)|-(twoch.east);
    \draw[-latex](twoch.south)--(twoch|-sum.north);
    
    \draw[-latex](m1.south)--(absphim1);
    \draw[-latex](absphim1.south)--(absphim1|-two1.north);
    \draw[-latex](m1.south)|-+(0.75,-0.05)|-(two1.east);
    \draw[-latex](two1.south)--(two1|-sum.north);
    \draw[-latex](sum.south)--+(0,-0.15)-|(minus1.north);
    \begin{scope}[on background layer]
    \draw[-latex](two1.south)|-+(0.75, -0.15)|-(minus1.east)node[xshift=0.15*\unit, yshift=0.15*\unit] {-};
    \end{scope}
    \draw[-latex](minus1.south)--(sm1.north);
    \draw[-latex](sm1.south)--(quant1.north);
    \draw[-latex](quant1.south)--(mv1.north);
    \draw[-latex](sm1.south)|-+(-0.5,-0.15)|-+(0,-1.6)-|(mv1.north);
    
    \draw[-latex](m2.south)--(absphim2);
    \draw[-latex](absphim2.south)--(absphim2|-two2.north);
    \draw[-latex](m2.south)|-+(0.75,-0.05)|-(two2.east);
    \draw[-latex](two2.south)--(two2|-sum.north);
    \draw[-latex](sum.south)--+(0,-0.15)-|(minus2.north);
    \begin{scope}[on background layer]
    \draw[-latex](two2.south)|-+(0.75, -0.15)|-(minus2.east)node[xshift=0.15*\unit, yshift=0.15*\unit] {-};
    \end{scope}
    \draw[-latex](minus2.south)--(sm2.north);
    \draw[-latex](sm2.south)--(quant2.north);
    \draw[-latex](quant2.south)--(mv2.north);
    \draw[-latex](sm2.south)|-+(-0.5,-0.15)|-+(0,-1.6)-|(mv2.north);
    
    \draw[-latex](mdv.south)--(absphimdv);
    \draw[-latex](absphimdv.south)--(absphimdv|-twodv.north);
    \draw[-latex](mdv.south)|-+(0.75,-0.05)|-(twodv.east);
    \draw[-latex](twodv.south)--(twodv|-sum.north);
    \draw[-latex](sum.south)--+(0,-0.15)-|(minusdv.north);
    \begin{scope}[on background layer]
    \draw[-latex](twodv.south)|-+(0.75, -0.15)|-(minusdv.east)node[xshift=0.15*\unit, yshift=0.15*\unit] {-};
    \end{scope}
    \draw[-latex](minusdv.south)--(smdv.north);
    \draw[-latex](smdv.south)--(quantdv.north);
    \draw[-latex](quantdv.south)--(mvdv.north);
    \draw[-latex](smdv.south)|-+(-0.5,-0.15)|-+(0,-1.6)-|(mvdv.north);

    \node (labelquant1input) at (quant1)[red, xshift=0.3*\unit, yshift=0.6*\unit]{$|y|$};
    \node (labelquant1input) at (quant1)[xshift=-1.05*\unit, yshift=0.6*\unit]{\footnotesize$\operatorname{sgn}(y)$};
    \node (labelquant1output) at (quant1)[red, xshift=0.3*\unit, yshift=-0.6*\unit]{$|t|$};
    \node (labelquant1output) at (quant1)[xshift=-1.05*\unit, yshift=-0.6*\unit]{\footnotesize$\operatorname{sgn}(t)$};
    
    \end{tikzpicture}
	\caption{A full variable node update\cite{mohr_uniform_2022}. The internal summation operation uses the two's complement (2's) format. The exchanged messages use the sign-magnitude (SM) format.%
	}
	\label{fig:full_vnu_hardware}
\end{figure}
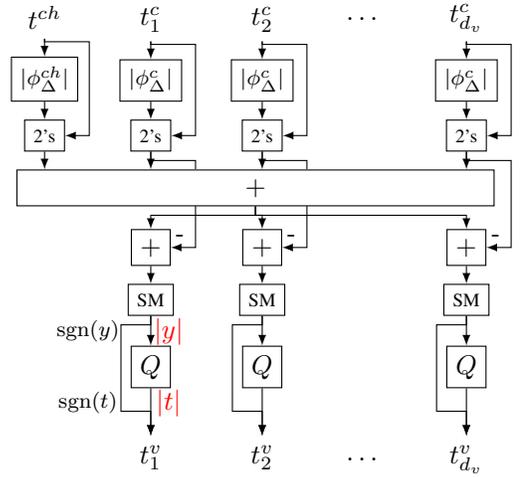
\begin{figure}
	\centering
	    \begin{tikzpicture} [scale=1.0, x=\unit,y=\unit, node distance=0.5*\unit and 0.5*\unit]   
     \node(y)[yshift=0*\unit, label={[yshift=-0.7*\unit, xshift=-0.35*\unit, red]{$|y|$}}]{};
     \node(orclip)[or gate US, draw, logic gate inputs=nnn, below=of y, yshift=-0.7*\unit, rotate=-90]{};
     
     \node(orclip1)[or gate US, draw, logic gate inputs=nn, below=of y, xshift=1.0*\unit, yshift=-2.3*\unit, rotate=-90]{};
     \node(orclip2)[or gate US, draw, logic gate inputs=nn, right of=orclip1,xshift=0.4*\unit, rotate=-90]{};
     \node(orclip3)[or gate US, draw, logic gate inputs=nn, right of=orclip2,xshift=0.4*\unit, rotate=-90]{};
     
     \node(deadend1)[circle,fill=black,inner sep=0pt,minimum size=4pt, right of= orclip, xshift=3*\unit]{};
     \node(deadend2)[circle,fill=black,inner sep=0pt,minimum size=4pt, right of= deadend1]{};
     
     \node(t)[below of=orclip3, yshift=-1.0*\unit, label={[yshift=-0.3*\unit, xshift=-0.4*\unit, red]{$|t|$}}]{};
     
     \draw[-latex](y)|-+(0,-0.5)-|(orclip.input 1);
     \draw[-latex](y)|-+(0,-0.5)-|(orclip.input 2);
     \draw[-latex](y)|-+(0,-0.5)-|(orclip.input 3);
     
     \draw[-latex](orclip.east)|-+(0,-0.15)-|(orclip1.input 2);
     \draw[-latex](y)|-+(0,-0.5)-|(orclip1.input 1);
     
     \draw[-latex](orclip.east)|-+(0,-0.15)-|(orclip2.input 2);
     \draw[-latex](y)|-+(0,-0.5)-|(orclip2.input 1);
    
     \draw[-latex](orclip.east)|-+(0.0,-0.15)-|(orclip3.input 2);
     \draw[-latex](y)|-+(0,-0.5)-|(orclip3.input 1);
     
     \draw[-latex](y)|-+(0,-0.5)-|(deadend1);
     \draw[-latex](y)|-+(0,-0.5)-|(deadend2);
     
     \draw[-latex](orclip1.east)|-+(0,-0.15)-|(t.north);
     \draw[-latex](orclip2.east)|-+(0,-0.15)-|(t.north);
     \draw[-latex](orclip3.east)|-+(0,-0.15)-|(t.north);
   
     \coordinate (ybitwidth0) at ($(y.south)+(0, -0.15*\unit)$);
     \draw ($(ybitwidth0)+(0.1*\unit, 0)$)--($(ybitwidth0)-(0.1*\unit,0)$)node[xshift=0.3*\unit]{\scriptsize$8$};
     
     \coordinate (tbitwidth) at ($(t.north)-(0,-0.3*\unit)$);
     \draw ($(tbitwidth)+(0.1*\unit, 0)$)--($(tbitwidth)-(0.1*\unit,0)$)node[xshift=0.3*\unit]{\scriptsize$3$};
     
     \node (MSB) at ($(orclip.input 1)+(-1.0*\unit, 0.3*\unit)$){\footnotesize MSB};
     \node (LSB) at ($(deadend2.west|-MSB)+(0.5*\unit,0)$){\footnotesize LSB};
    
     \node (rshift) at ($(deadend1)+(0.3*\unit, -0.4*\unit)$){$r=2$};

    \end{tikzpicture}
	\caption{Hardware schematics from \cite{mohr_uniform_2022} for performing uniform quantization with $w_y{=}9\,$bit for $y$ and $w{=}4\,$bit for $t$.}
	\label{fig:hardware_quant_uniform}
\end{figure}
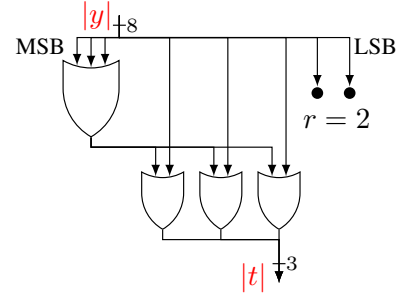

\section{Design of Check Node Aware Quantization}
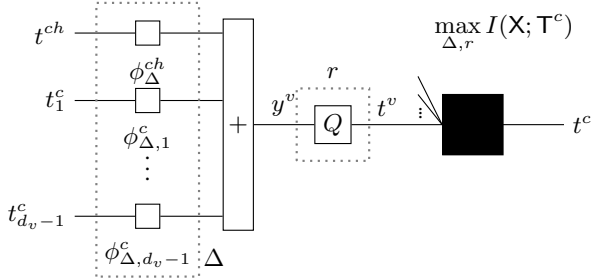
\begin{figure}[t]
    \centering
    \small
\begin{tikzpicture} [scale=1.0, x=\unit,y=\unit, varnode/.style={circle, inner sep=0pt, minimum height=1.0*\unit, draw=black},  checknode/.style={rectangle, inner sep=0pt, minimum width=0.8*\unit,minimum height=0.8*\unit,draw=black, fill=black}, node distance=1.5*\unit and 1.0*\unit]

\node(phich)[rectangle, minimum width=0.4*\unit,minimum height=0.4*\unit,draw=black, fill=white,label={[shift={(0.0,-1.2)}]$\phi_\Delta^{ch}$}]{};
\node(ych)[left=of phich]{$t^{ch}$};
\node(phic1)[rectangle, below=of phich, yshift=0.8*\unit,minimum width=0.4*\unit,minimum height=0.4*\unit,draw=black, fill=white,label={[shift={(0.0,-1.2)}]$\phi^{c}_{\Delta,1}$}]{};
\node(yc1)[left=of phic1]{$t^c_1$};
\node(dotsphi1tophi2)[below=of phic1, yshift=1.2*\unit]{$\vdots$};
\node(phic2)[rectangle, below=of phic1, minimum width=0.4*\unit,minimum height=0.4*\unit,draw=black, fill=white,label={[shift={(0.0,-1.2)}]$\phi^{c}_{\Delta,d_v-1}$}]{};
\node(yc2)[left=of phic2]{$t^c_{d_v-1}$};

\node(sum) at ($(phich)!0.5!(phic2)$)[rectangle, xshift=1.5*\unit, inner sep=0pt,minimum width=0.5*\unit,minimum height=3.5*\unit,draw=black, fill=white]{$+$};

\node(qv)[rectangle, right=of sum, minimum width=0.5*\unit,minimum height=0.5*\unit,draw=black, fill=white]{$Q$};

\node(cn)[rectangle, right=of qv, xshift=0.5*\unit, inner sep=0pt,minimum width=1.0*\unit,minimum height=1.0*\unit,draw=black, fill=black]{};
\node(tc)[right=of cn]{$t^c$};

\draw(ych)--(phich)--(sum.west|-phich);
\draw(yc1)--(phic1)--(sum.west|-phic1);
\draw(yc2)--(phic2)--(sum.west|-phic2);
\draw(sum.east)--node[above, midway]{$y^{v}$}(qv.west);
\draw(qv.east)--node[above, pos=0.4]{$t^{v}$}(cn.west)--(tc);
\draw(cn.west)--($(cn.west)+(-0.4*\unit,0.8*\unit)$);
\draw(cn.west)--node[below, scale=0.4, xshift=-0.4*\unit, yshift=0.5*\unit]{\huge$\vdots$}($(cn.west)+(-0.4*\unit,0.5*\unit)$);

\draw[gray,thick,dotted] ($(phic2)+(-0.85*\unit,-1.0*\unit)$)  rectangle ($(phich)+(0.85*\unit,0.5*\unit)$);
\draw[gray,thick,dotted] ($(qv)+(-0.6*\unit,-0.6*\unit)$)  rectangle ($(qv)+(0.6*\unit,0.6*\unit)$);

\node[below=of phic2, yshift=1.3*\unit, xshift=1.1*\unit]{$\Delta$};
\node[above=of qv, yshift=-1.2*\unit]{$r$};
\node(maxmi) at ($(tc)+(-1.3*\unit,1.5*\unit)$){$\begin{aligned}
    \max_{\Delta, r} I(\mathsf{X};\mathsf{T}^c)
    \end{aligned}$};

\end{tikzpicture}
    \vspace{-0.1cm}
    \caption{Setup with variable and check node.}
    \label{fig:awarevsnotawaresetup}
\end{figure}

We consider a check node where extrinsic information about some code bit $x$ is obtained from $x=b_1\oplus \ldots \oplus b_{d_c-1}$.
Typically, the $d_c{-}1$ connected variable nodes are designed to preserve extrinsic information about the underlying code bit $b_i$ in the corresponding variable node message $t^v_i$. All the variable nodes within one iteration use the same design obtained for a template variable node $b$ and output message $t^v$ that is typically optimized for $\max_{Q^v} I(\myvar{B};\myvar{T}^v)$\cite{lewandowsky_trellis_2015,he_mutual_2019}. 
However, as we will show, this local optimization does not achieve the maximum mutual information $I(\myvar{X};\myvar{T}^c)$ between the code bit $x$ and the check node message $t^c$. Therefore, we extend the local optimization of the variable node by taking the check node behavior into account, aiming for $\max_{Q^v}I(\myvar{X};\myvar{T}^c)$.
Note that with restriction to uniform boundaries, $Q^v$ is defined by $\Delta$ and $r$ as depicted in Fig.~\ref{fig:awarevsnotawaresetup}\cite{mohr_uniform_2022}. To simplify computations, we assume the minimum approximation (\ref{equ:cn_min_approx}) in the check node. Appendix \ref{sec:optimization_procedure} describes the optimization procedure in detail.

\subsection{Example with 2-Bit Decoding}\label{sec:example}
\begin{table}[t]
    \centering
    \caption{Translation tables used in Section \ref{sec:example}.}%
    \label{fig:translation_tables}
    \setlength{\tabcolsep}{3pt}
    \begin{tabular}{c|cccccccc}
        $t$ & +1 & +2 & +3 & +4 & +5 & +6& +7& +8\\
        \hline
        $\phi^{ch}(t)$& .30 & .93 & 1.60 & 2.36 & 3.23 & 4.32 & 5.83 & 8.64 \\
        $\phi^{ch}_{\Delta}(t)$& 6 & 19 & 32 & 48 & 65 & 87 & 118 & 127
        
    \end{tabular}
    \begin{tabular}{c|cc}
        t & +1 & +2 \\
        \hline
        $\phi^{c}(t)$ & .17 & .98\\
        $\phi^{c}_{\Delta}(t)$ & 3 & 20\\
    \end{tabular}
    \vspace{-0.0cm}
\end{table}
\begin{figure}[t]
        \begin{subfigure}[b]{0.5\textwidth}
    
    \pgfkeys{/includestandaloneresized, default, width=0.75\textwidth, ratio=0.65}
    \setlength\figurewidth{\itgWidth}
    \setlength\figureheight{\itgRatio\figurewidth}
    \includestandalone{standalones/mi_vtc_vs_mi_ctv/source}

    \vspace{-0.1cm}
    \caption{Optimization of mutual information before and after the check node.}
    \vspace{0.2cm}
    \label{fig:mi_vtc_vs_mi_ctv}
\end{subfigure}
    \begin{subfigure}[b]{0.46\textwidth}
        \hspace{-0.6cm}
        
    \pgfkeys{/includestandaloneresized, default, width=0.95\textwidth, ratio=0.6}
    \setlength\figurewidth{\itgWidth}
    \setlength\figureheight{\itgRatio\figurewidth}
    \includestandalone{standalones/boundaries/source}

        \vspace{-0.6cm}
        \caption{Placement of quantizer boundaries.}
        \vspace{0.2cm}
        \label{fig:boundaries}
    \end{subfigure}
    \hfill
    \begin{subfigure}[b]{0.5\textwidth}
        \begin{subfigure}[b]{1.0\linewidth}
            \centering
            
    \pgfkeys{/includestandaloneresized, default, width=1.0\textwidth, ratio=0.5}
    \setlength\figurewidth{\itgWidth}
    \setlength\figureheight{\itgRatio\figurewidth}
    \includestandalone{standalones/ctv_vtc_legend/source}

            \vspace{-0.5cm}
            \label{fig:ctv_vtc_legend}
        \end{subfigure}
        \hspace*{-0.3cm}
        \begin{subfigure}[b]{0.22\linewidth}
            
    \pgfkeys{/includestandaloneresized, default, width=1.25\textwidth, ratio=1.8}
    \setlength\figurewidth{\itgWidth}
    \setlength\figureheight{\itgRatio\figurewidth}
    \pgfplotsset{legend style={nodes={scale=0.7, transform shape}}}
\pgfplotsset{ every non boxed x axis/.append style={x axis line style=-},
               every non boxed y axis/.append style={y axis line style=-}}
\begin{tikzpicture}

\definecolor{color0}{rgb}{0,0.270588235294118,0.541176470588235}

\begin{axis}[
height=\figureheight,
tick align=outside,
tick pos=left,
width=\figurewidth,
x grid style={white!69.0196078431373!black},
xlabel={\(\displaystyle t^v\)},
xmin=1.6, xmax=3.4,
xtick={2,3},
xticklabels={+1,+2},
xtick style={color=black},
y grid style={white!69.0196078431373!black},
ylabel shift = -0.2cm,
ylabel={\(\displaystyle p(t^v)\)},
ymin=0, ymax=0.501861564913527,
ytick style={color=black},
ytick={0,0.2,0.4,0.6},
yticklabels={0.0,0.2,0.4,0.6}
]
\draw[draw=none,fill=color0] (axis cs:-0.4,0) rectangle (axis cs:0,0.477963395155739);
\draw[draw=none,fill=color0] (axis cs:0.6,0) rectangle (axis cs:1,0.0220366048442604);
\draw[draw=none,fill=color0] (axis cs:1.6,0) rectangle (axis cs:2,0.0220366048442604);
\draw[draw=none,fill=color0] (axis cs:2.6,0) rectangle (axis cs:3,0.47796339515574);
\draw[draw=none,fill=red!75.2941176470588!black] (axis cs:-2.77555756156289e-17,0) rectangle (axis cs:0.4,0.453100254688678);
\draw[draw=none,fill=red!75.2941176470588!black] (axis cs:1,0) rectangle (axis cs:1.4,0.0468997453113214);
\draw[draw=none,fill=red!75.2941176470588!black] (axis cs:2,0) rectangle (axis cs:2.4,0.0468997453113215);
\draw[draw=none,fill=red!75.2941176470588!black] (axis cs:3,0) rectangle (axis cs:3.4,0.453100254688679);
\end{axis}

\end{tikzpicture}

            \vspace{-0.5cm}
            \label{fig:vtc_1}
        \end{subfigure}
        \begin{subfigure}[b]{0.22\linewidth}
            
    \pgfkeys{/includestandaloneresized, default, width=1.25\textwidth, ratio=1.8}
    \setlength\figurewidth{\itgWidth}
    \setlength\figureheight{\itgRatio\figurewidth}
    \pgfplotsset{legend style={nodes={scale=0.7, transform shape}}}
\pgfplotsset{ every non boxed x axis/.append style={x axis line style=-},
               every non boxed y axis/.append style={y axis line style=-}}
\begin{tikzpicture}

\definecolor{color0}{rgb}{0,0.270588235294118,0.541176470588235}

\begin{axis}[
height=\figureheight,
tick align=outside,
tick pos=left,
width=\figurewidth,
x grid style={white!69.0196078431373!black},
xlabel={\(\displaystyle t^c\)},
xmin=1.6, xmax=3.4,
xtick style={color=black},
xtick={2,3},
xticklabels={+1,+2},
xtick style={color=black},
y grid style={white!69.0196078431373!black},
ylabel shift = -0.2cm,
ylabel={\(\displaystyle p(t^c)\)},
ymin=0, ymax=0.500219351688643,
ytick style={color=black},
ytick={0,0.2,0.4,0.6},
yticklabels={0.0,0.2,0.4,0.6}
]
\draw[draw=none,fill=color0] (axis cs:-0.4,0) rectangle (axis cs:0,0.123632778654208);
\draw[draw=none,fill=color0] (axis cs:0.6,0) rectangle (axis cs:1,0.376367221345791);
\draw[draw=none,fill=color0] (axis cs:1.6,0) rectangle (axis cs:2,0.376367221345791);
\draw[draw=none,fill=color0] (axis cs:2.6,0) rectangle (axis cs:3,0.123632778654208);
\draw[draw=none,fill=red!75.2941176470588!black] (axis cs:-2.77555756156289e-17,0) rectangle (axis cs:0.4,0.0236006174393857);
\draw[draw=none,fill=red!75.2941176470588!black] (axis cs:1,0) rectangle (axis cs:1.4,0.476399382560612);
\draw[draw=none,fill=red!75.2941176470588!black] (axis cs:2,0) rectangle (axis cs:2.4,0.476399382560612);
\draw[draw=none,fill=red!75.2941176470588!black] (axis cs:3,0) rectangle (axis cs:3.4,0.0236006174393857);
\end{axis}

\end{tikzpicture}

            \vspace{-0.5cm}
            \label{fig:vtc_1}
        \end{subfigure}
        \begin{subfigure}[b]{0.22\linewidth}
            
    \pgfkeys{/includestandaloneresized, default, width=1.25\textwidth, ratio=1.8}
    \setlength\figurewidth{\itgWidth}
    \setlength\figureheight{\itgRatio\figurewidth}
    \pgfplotsset{legend style={nodes={scale=0.7, transform shape}}}
\pgfplotsset{ every non boxed x axis/.append style={x axis line style=-},
               every non boxed y axis/.append style={y axis line style=-}}
\begin{tikzpicture}

\definecolor{color0}{rgb}{0,0.270588235294118,0.541176470588235}

\begin{axis}[
height=\figureheight,
tick align=outside,
tick pos=left,
width=\figurewidth,
x grid style={white!69.0196078431373!black},
xlabel={\(\displaystyle t^c\)},
xmin=1.6, xmax=3.4,
xtick style={color=black},
xtick={2,3},
xticklabels={+1,+2},
xtick style={color=black},
y grid style={white!69.0196078431373!black},
ylabel shift = -0.2cm,
ylabel={\(\displaystyle D_{{KL}}(p(x|t^c)||p(x))\)},
ymin=0, ymax=0.550586257181518,
ytick style={color=black},
ytick={0,0.2,0.4,0.6},
yticklabels={0.0,0.2,0.4,0.6}
]
\draw[draw=none,fill=color0] (axis cs:-0.4,0) rectangle (axis cs:0,0.257273734787868);
\draw[draw=none,fill=color0] (axis cs:0.6,0) rectangle (axis cs:1,0.0126187071953285);
\draw[draw=none,fill=color0] (axis cs:1.6,0) rectangle (axis cs:2,0.0126187071953286);
\draw[draw=none,fill=color0] (axis cs:2.6,0) rectangle (axis cs:3,0.257273734787868);
\draw[draw=none,fill=red!75.2941176470588!black] (axis cs:-2.77555756156289e-17,0) rectangle (axis cs:0.4,0.524367863982399);
\draw[draw=none,fill=red!75.2941176470588!black] (axis cs:1,0) rectangle (axis cs:1.4,0.0336333921556034);
\draw[draw=none,fill=red!75.2941176470588!black] (axis cs:2,0) rectangle (axis cs:2.4,0.0336333921556034);
\draw[draw=none,fill=red!75.2941176470588!black] (axis cs:3,0) rectangle (axis cs:3.4,0.524367863982399);
\end{axis}

\end{tikzpicture}

            \vspace{-0.5cm}
            \label{fig:vtc_1}
        \end{subfigure}
        \begin{subfigure}[b]{0.22\linewidth}
            
    \pgfkeys{/includestandaloneresized, default, width=1.25\textwidth, ratio=1.8}
    \setlength\figurewidth{\itgWidth}
    \setlength\figureheight{\itgRatio\figurewidth}
    \pgfplotsset{legend style={nodes={scale=0.7, transform shape}}}
\pgfplotsset{ every non boxed x axis/.append style={x axis line style=-},
               every non boxed y axis/.append style={y axis line style=-}}
\begin{tikzpicture}

\definecolor{color0}{rgb}{0,0.270588235294118,0.541176470588235}

\begin{axis}[
height=\figureheight,
tick align=outside,
tick pos=left,
width=\figurewidth,
x grid style={white!69.0196078431373!black},
xlabel={\(\displaystyle t^c\)},
xmin=1.6, xmax=3.4,
xtick style={color=black},
xtick={2,3},
xticklabels={+1,+2},
xtick style={color=black},
y grid style={white!69.0196078431373!black},
ylabel shift = -0.2cm,
ylabel={\(\displaystyle p(t^c)D_{{KL}}(p(x|t^c)||p(x))\)},
ymin=0, ymax=0.0333978400418985,
ytick style={color=black}
]
\draw[draw=none,fill=color0] (axis cs:-0.4,0) rectangle (axis cs:0,0.03180746670657);
\draw[draw=none,fill=color0] (axis cs:0.6,0) rectangle (axis cs:1,0.00474926776408192);
\draw[draw=none,fill=color0] (axis cs:1.6,0) rectangle (axis cs:2,0.00474926776408197);
\draw[draw=none,fill=color0] (axis cs:2.6,0) rectangle (axis cs:3,0.03180746670657);
\draw[draw=none,fill=red!75.2941176470588!black] (axis cs:-2.77555756156289e-17,0) rectangle (axis cs:0.4,0.0123754053553564);
\draw[draw=none,fill=red!75.2941176470588!black] (axis cs:1,0) rectangle (axis cs:1.4,0.0160229272563484);
\draw[draw=none,fill=red!75.2941176470588!black] (axis cs:2,0) rectangle (axis cs:2.4,0.0160229272563484);
\draw[draw=none,fill=red!75.2941176470588!black] (axis cs:3,0) rectangle (axis cs:3.4,0.0123754053553564);
\end{axis}

\end{tikzpicture}

            \vspace{-0.5cm}
            \label{fig:vtc_1}
        \end{subfigure}
        \caption{Probability distribution and Kullback Leibler divergence of the messages. Our design enforces symmetric distributions such that $p(\myvar{X}{=}0,t){=}p(\myvar{X}{=}1,{-}t)$ for $t$ being $t^v$ or $t^c$. Therefore, only the two outcomes +1 and +2 are shown.}
        \label{fig:distribution_vtc_ctv}
    \end{subfigure}
    \caption{Comparison of the check node aware and unaware variable node design.}
    \vspace{0.0cm}
\end{figure}
In this example, we compare two optimization procedures. The first option (proposed) is the check node aware design, where the mutual information in the messages \emph{after} the check node update shall be maximized. The second option (conventional) performs the mutual information maximization in the messages \emph{before} the check node update. The messages exchanged between the variable nodes ($d_v=6$) and check nodes ($d_c=32$) use 2~bits. 
We assume a variable node with given input joint distributions $p(x,t^c)$ and $p(x,t^{ch})$. Those distributions determine the translation tables $\phi^{ch}$ and $\phi^{c}$ shown in the Table~\ref{fig:translation_tables}. 

The results from the optimization search are shown in Fig.~\ref{fig:mi_vtc_vs_mi_ctv}.  
The two vertical lines depict the optimal boundary spacing for the two optimization criteria. Clearly, the check node aware solution leads to a higher mutual information $I(\myvar{X};\myvar{T^c})$ in the check-to-variable node messages, despite lower mutual information $I(\myvar{B};\myvar{T^v})$ in the variable-to-check node messages. 
The results indicate that the objective function $f(\Delta, r){=}I(\myvar{X};\myvar{T}^c)$ is concave (except for a few small outliers).

In Fig.~\ref{fig:boundaries}, the two found boundaries are depicted w.r.t. to the LLRs $L(b|y^v)$ and distribution $p(y^v)$ before quantization. Note, that the check node aware design places the boundary significantly closer to the decision threshold, which potentially reduces the dynamic range that must be covered by the adder unit in a hardware implementation (see Fig.~\ref{fig:full_vnu_hardware}). Furthermore, the optimization procedure improves the accuracy of small LLR magnitude representations by choosing a small value for~$\Delta$. This comes at the price of distortions for high LLR magnitudes caused by clipping effects that occur for $t^{ch}{=}+8$ in the translation table (see Table~\ref{fig:translation_tables}). However, the distortion for $L(b|y^v)>4.0$ is irrelevant for the performance since the threshold is placed at $L(b|y^v)\approx1.6$.

As described in Section~\ref{sec:introduction}, low-reliability messages dominate high-reliability messages in the check node update.
This behavior of the check node causes the check node aware design to decrease the frequency of low-reliability messages and to increase the probability for high-reliability messages, as can be observed in Fig.~\ref{fig:distribution_vtc_ctv}. In this way, the probability for low-reliability check node messages $t^c$ is reduced. On the other hand, the Kullback Leibler divergence $D_{\mylabel{KL}}(p(x|t^c)||p(x))$ is reduced for the high-reliability check node messages $t^c$. The proposed method finds a trade-off, which ultimately leads to a higher preservation of mutual information $I(\myvar{X};\myvar{T}^c)=\sum_{t^c} p(t^c) D_{\mylabel{KL}}(p(x|t^c)||p(x))$ in the check node messages. 

\begin{figure}[t]
    \hspace{-1.0cm}
    
    \pgfkeys{/includestandaloneresized, default, width=0.9\linewidth, ratio=0.75}
    \setlength\figurewidth{\itgWidth}
    \setlength\figureheight{\itgRatio\figurewidth}
        
    \pgfplotsset{legend style={nodes={scale=0.7, transform shape}}}
    \pgfplotsset{ every non boxed x axis/.append style={x axis line style=-},
        every non boxed y axis/.append style={y axis line style=-}}
    \pgfplotsset{
        compat=1.11,
        legend image code/.code={
            \draw[mark repeat=2,mark phase=2]
            plot coordinates {
                (0cm,0cm)
                (0.15cm,0cm)        %
                (0.3cm,0cm)         %
            };%
        }
    }
    \begin{tikzpicture}
    
    \definecolor{color0}{rgb}{0,0.270588235294118,0.541176470588235}
    
    \begin{axis}[
    height=\figureheight,
    legend cell align={left},
    legend style={
        fill opacity=0.8,
        draw opacity=1,
        text opacity=1,
        at={(0.01,0.99)},
        anchor=north west,
        draw=white!80!black,
        inner sep=1pt
    },
    tick align=outside,
    tick pos=left,
    width=\figurewidth,
    x grid style={white!69.0196078431373!black},
    xlabel={Iteration},
    xmin=-1.5, xmax=31.5,
    xtick style={color=black},
    y grid style={white!69.0196078431373!black},
    ylabel={$I(\mathsf{B};\mathsf{Y}^{app})$ \ref{pgfplots:plot1}\quad
            $I(\mathsf{B};\mathsf{T}^v)$ \ref{pgfplots:plot2}},%
    ylabel shift = -0.2cm,
    ymin=0.860383340232775, ymax=1.00664841236988,
    ytick style={color=black}
    ]
    \addplot [semithick, color0]
    table {%
        0 0.897075826591057
        1 0.916541497463295
        2 0.928426494073068
        3 0.937485868203909
        4 0.946133047285962
        5 0.955427557155072
        6 0.966533764719676
        7 0.979833590035202
        8 0.993339558227681
        9 0.99971155562748
        10 0.999999993674815
        11 1.00000000000001
    };
    \label{pgfplots:plot1}
    \addlegendentry{$I(\mathsf{B};\mathsf{Y}^{{app}})$}
    \addplot [semithick, red!75.2941176470588!black, forget plot]
    table {%
        0 0.897075826591057
        1 0.911856198827859
        2 0.917498147304332
        3 0.9204460406494
        4 0.922591507890709
        5 0.924110580347414
        6 0.926048094066106
        7 0.927657010311992
        8 0.929008582357322
        9 0.930262475792115
        10 0.931467068753062
        11 0.932447014756047
        12 0.933437395329447
        13 0.934444851678711
        14 0.935438976415126
        15 0.936439434633381
        16 0.937494464035695
        17 0.938599291864774
        18 0.939773873974034
        19 0.941099586083023
        20 0.94236835356815
        21 0.943697202854177
        22 0.945246720688631
        23 0.947379798213981
        24 0.950012712656556
        25 0.953259996141296
        26 0.9601348667223
        27 0.967154273222301
        28 0.978454733947902
        29 0.990888321298498
        30 0.999179444680253
    };
    \addplot [semithick, color0, dash pattern=on 1pt off 1pt on 1pt off 1pt]
    table {%
        0 0.867031752602644
        1 0.891116389132226
        2 0.901333824750426
        3 0.911616994148941
        4 0.921822694288528
        5 0.933372626532116
        6 0.946581354697452
        7 0.964807281466873
        8 0.986189566020153
        9 0.998979878774686
        10 0.999999861912293
        11 1
    };
    \label{pgfplots:plot2}
    \addlegendentry{$I(\mathsf{B};\mathsf{T}^{{v}})$}
    \addplot [semithick, red!75.2941176470588!black, dash pattern=on 1pt off 1pt on 1pt off 1pt, forget plot]
    table {%
        0 0.882130829114794
        1 0.894980762616216
        2 0.900511783007718
        3 0.903410357228412
        4 0.905507581447249
        5 0.907055440146883
        6 0.908956716398352
        7 0.910584157044153
        8 0.911950033278437
        9 0.913146474448743
        10 0.914440886224523
        11 0.915506415840425
        12 0.916532787822931
        13 0.917553012136667
        14 0.918582441316911
        15 0.919634561231235
        16 0.920723534537606
        17 0.921864949525842
        18 0.923076916214374
        19 0.924386931293086
        20 0.925651365111452
        21 0.927036524771912
        22 0.928594177748705
        23 0.930816513711953
        24 0.9335570042072
        25 0.936969516046075
        26 0.944616490050166
        27 0.952373768058045
        28 0.966138698972822
        29 0.983218182545841
        30 0.997625915955821
    };
    \legend{}; %
    \end{axis}
    
    \begin{axis}[
    axis y line=right,
    height=\figureheight,
    legend cell align={left},
    legend style={fill opacity=0.8, draw opacity=1, text opacity=1, draw=white!80!black, at={(0.99,0.99)}, inner sep=1pt},
    tick align=outside,
    width=\figurewidth,
    x grid style={white!69.0196078431373!black},
    xmin=-1.5, xmax=31.5,
    xtick pos=left,
    xtick style={color=black},
    y grid style={white!69.0196078431373!black},
    ylabel={$I(\mathsf{X};\mathsf{T}^c)$ \ref{pgfplots:plot3}},%
    ylabel shift = -0.2cm,
    ymin=-0.0163589755475848, ymax=1.04839375815303,
    ytick pos=right,
    ytick style={color=black},
    yticklabel style={anchor=west}
    ]
    \addplot [semithick, color0, dashed]
    table {%
        0 0.0441095716121332
        1 0.0731134689413039
        2 0.0990801706787
        3 0.126587101512846
        4 0.160482684849182
        5 0.208891206004462
        6 0.28878518116996
        7 0.436263813663216
        8 0.708824519275705
        9 0.972458040588809
        10 0.999995906621186
    };
    \label{pgfplots:plot3}
    \addlegendentry{$I(\mathsf{X};\mathsf{T}^{{c}})$}
    \addplot [semithick, red!75.2941176470588!black, dashed, forget plot]
    table {%
        0 0.0320388759842614
        1 0.0448765300947609
        2 0.0521501002719951
        3 0.0574377729276484
        4 0.0613466035648461
        5 0.0662473659183637
        6 0.0705150489758773
        7 0.0741779495315551
        8 0.0773724885535345
        9 0.0807740297343069
        10 0.0835064701366465
        11 0.0862421171454724
        12 0.089014577999797
        13 0.0918619324161255
        14 0.0948234610370795
        15 0.0979440913220823
        16 0.101276442195845
        17 0.104884401973925
        18 0.10884845712699
        19 0.112802248946338
        20 0.117116892349675
        21 0.122084989463183
        22 0.129140620052514
        23 0.138258154971587
        24 0.1500994063648
        25 0.176875319235628
        26 0.209042596118839
        27 0.273996472963579
        28 0.3907780652457
        29 0.625950983770948
    };
    \node at (2, 0.7)[color0]{aware};
    \node at (15, 0.6)[red!75.2941176470588!black]{unaware};
    \legend{}; %
    \end{axis}
    \end{tikzpicture}
    

    \vspace{-0.4cm}
    \caption{Evolution of mutual information for 2 bits and $E_b/N_0=3.45\,$dB. The message $y^{app}$ includes the non-extrinsic information in the variable node update.}
    \label{fig:dde_mi_app_mi_ctv}
\end{figure}
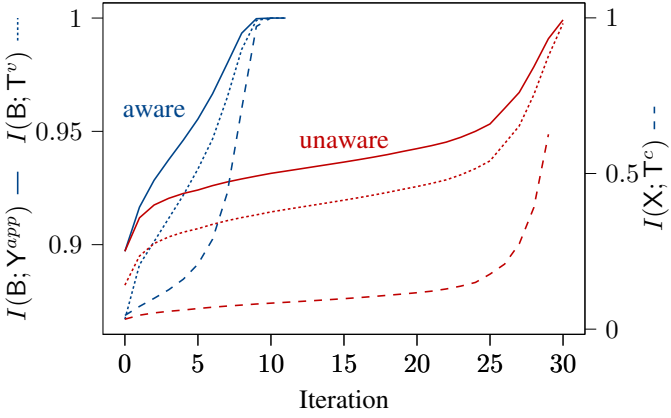

The overall mutual information gain of the check node aware design is $\Delta I(\myvar{X},\myvar{T}^c)=0.073-0.056=0.017$. The gains accumulate over multiple iterations as shown in Fig.~\ref{fig:dde_mi_app_mi_ctv}. Remarkably, the check node aware design requires only one third of the iteration count for convergence in the considered scenario. Note, that the initial mutual information for the variable-to-check node messages is lower in the beginning. However, after a few iterations, the mutual information $I(\myvar{B},\myvar{T}^v)$ of the check node unaware design is surpassed.

\section{Performance Evaluation}
To evaluate the error rate performance, we first consider a regular high-rate LDPC code \cite{noauthor_standard_2006} with $d_v{=}6$, $d_c{=}32$, $N{=}2048$ and rate $R{=}0.84$ (the parity check matrix does not have full rank). The decoder performs 10 iterations.

\begin{figure}[t]
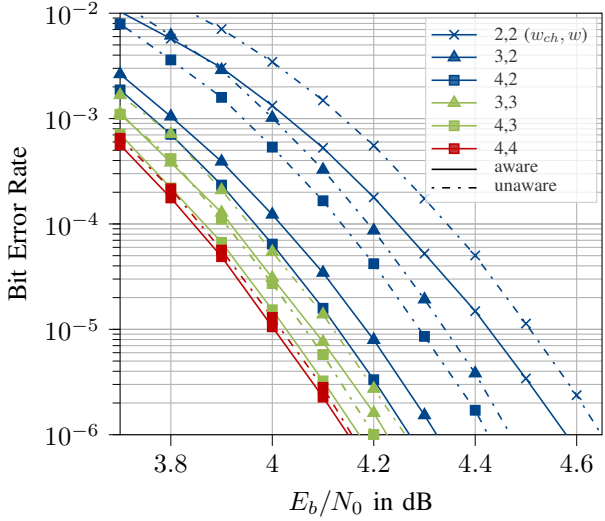

    \vspace{-0.4cm}
    
    \pgfkeys{/includestandaloneresized, default, width=0.9\linewidth, ratio=0.9}
    \setlength\figurewidth{\itgWidth}
    \setlength\figureheight{\itgRatio\figurewidth}
    \includestandalone{standalones/code3_10_ber_2bit_3bit_4bit_medium_ber/source}

    \vspace{-0.1cm}
    \caption{Bit error rates for code with rate $R{=}0.84$. The numbers specify the channel and decoder message bit width, $w_{ch}$ and $w$, respectively. %
    }
    \label{fig:code3_10_ber_2bit_3bit_4bit}
\end{figure}

In Fig.~\ref{fig:code3_10_ber_2bit_3bit_4bit} the bit error rates for the check node aware and unaware design are shown. In all cases the minimum approximation in the check node and computational domain with uniform quantization in the variable node are used. The design $E_b/N_0$ is optimized for best performance at bit error rates equal to $10^{-6}$.
As expected from Fig.~\ref{fig:dde_mi_app_mi_ctv}, the check node aware designs also show superior error rate performance over the check node unaware configurations. Especially, the 2-bit decoder with 3-bit (4-bit) channel messages achieves a gain of 0.16\,dB (0.16\,dB). With increasing decoder bit width $w$ the gains reduce to 0.05\,dB for 3-bit decoding and to 0.02\,dB for 4-bit decoding.

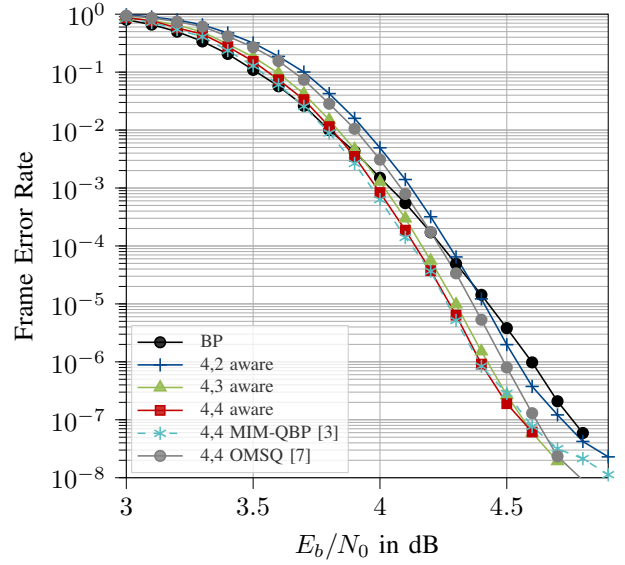
\begin{figure}[t]
    
    \pgfkeys{/includestandaloneresized, default, width=0.9\linewidth, ratio=0.97}
    \setlength\figurewidth{\itgWidth}
    \setlength\figureheight{\itgRatio\figurewidth}
    \pgfplotsset{legend style={nodes={scale=0.7, transform shape}}}
\pgfplotsset{ every non boxed x axis/.append style={x axis line style=-},
               every non boxed y axis/.append style={y axis line style=-}}
\begin{tikzpicture}

\definecolor{color0}{rgb}{0,0.270588235294118,0.541176470588235}
\definecolor{color1}{rgb}{0.584313725490196,0.729411764705882,0.309803921568627}
\definecolor{color2}{rgb}{0.341176470588235,0.741176470588235,0.764705882352941}

\begin{axis}[
height=\figureheight,
legend cell align={left},
legend style={
  fill opacity=0.8,
  draw opacity=1,
  text opacity=1,
  inner sep=2pt,
  at={(0.01,0.01)},
  anchor=south west,
  draw=white!80!black
},
log basis y={10},
tick align=outside,
tick pos=left,
width=\figurewidth,
x grid style={white!69.0196078431373!black},
xlabel={\(\displaystyle E_b/N_0\) in dB},
xmajorgrids,
xmin=3, xmax=4.9,
xminorgrids,
xtick style={color=black},
y grid style={white!69.0196078431373!black},
ymajorgrids,
ymin=1e-08, ymax=1,
yminorgrids,
ymode=log,
ytick style={color=black},
ylabel={Frame Error Rate}
]
\addplot [semithick, black, mark=*, mark size=2, mark options={solid}]
table {%
2 1
2.1 1
2.2 1
2.3 1
2.4 1
2.5 0.9995
2.6 0.998
2.7 0.9895
2.8 0.9665
2.9 0.9165
3 0.8015
3.1 0.661
3.2 0.50025
3.3 0.33875
3.4 0.206
3.5 0.1095
3.6 0.057
3.7 0.02625
3.8 0.01
3.9 0.00426050420168067
4 0.00150149700598802
4.1 0.000547645125958379
4.2 0.000171291538198013
4.3 4.91014435824413e-05
4.4 1.43299323627192e-05
4.5 3.81743498908214e-06
4.6 9.781921834619e-07
4.7 2.08685640269133e-07
4.8 5.90841949778434e-08
};
\addlegendentry{BP}
\addplot [semithick, color0, mark=+, mark size=2.5, mark options={solid}]
table {%
2 1
2.1 1
2.2 1
2.3 1
2.4 1
2.5 1
2.6 0.999
2.7 0.999
2.8 0.992
2.9 0.978
3 0.976
3.1 0.8955
3.2 0.794
3.3 0.6555
3.4 0.474
3.5 0.31875
3.6 0.187
3.7 0.1004
3.8 0.0425833333333333
3.9 0.0159375
4 0.00494607843137255
4.1 0.00140196078431373
4.2 0.000319108280254777
4.3 6.46412411118293e-05
4.4 1.21986923001854e-05
4.5 1.9802450751305e-06
4.6 3.76090683578802e-07
4.7 1.21408120735707e-07
4.8 4.22735632841746e-08
4.9 2.29108487960349e-08
};
\addlegendentry{4,2 aware}
\addplot [semithick, color1, mark=triangle*, mark size=2.5, mark options={solid}]
table {%
2 1
2.1 1
2.2 1
2.3 1
2.4 1
2.5 0.9995
2.6 0.999
2.7 0.993
2.8 0.981
2.9 0.9545
3 0.8835
3.1 0.7785
3.2 0.651
3.3 0.49375
3.4 0.30575
3.5 0.1825
3.6 0.0956666666666667
3.7 0.0422083333333333
3.8 0.015
3.9 0.00466666666666667
4 0.00128865979381443
4.1 0.000297973778307509
4.2 5.58098002009153e-05
4.3 9.81238715754769e-06
4.4 1.51514233247071e-06
4.5 2.67068809375119e-07
4.6 5.99022317018322e-08
4.7 1.9288039872236e-08
};
\addlegendentry{4,3 aware}
\addplot [semithick, red!75.2941176470588!black, mark=square*, mark size=1.75, mark options={solid}]
table {%
3 0.868
3.1 0.7675
3.2 0.581
3.3 0.45075
3.4 0.2755
3.5 0.153625
3.6 0.0739285714285714
3.7 0.0335
3.8 0.0114659090909091
3.9 0.0035
4 0.000835559265442404
4.1 0.000187336080929187
4.2 3.7450501305923e-05
4.3 6.38406537282942e-06
4.4 9.13609121473469e-07
4.5 1.89426227955523e-07
4.6 6.15331293112434e-08
};
\addlegendentry{4,4 aware}
\addplot [semithick, color2, dashed, mark=asterisk, mark size=2.5, mark options={solid}]
table {%
3 0.8395
3.1 0.7225
3.2 0.5455
3.3 0.40775
3.4 0.235666666666667
3.5 0.128
3.6 0.0605
3.7 0.0256
3.8 0.00889473684210526
3.9 0.00265079365079365
4 0.000630352644836272
4.1 0.00014010989010989
4.2 3.66032210834553e-05
4.3 5.17812758906379e-06
4.4 8.40477391158178e-07
4.5 2.85936499222253e-07
4.6 7.68044435978888e-08
4.7 3.16457298517967e-08
4.8 2.13863303708988e-08
4.9 1.12719645501224e-08
};
\addlegendentry{4,4 MIM-QBP\cite{{he_mutual_2019}}}
\addplot [semithick, gray, mark=*, mark size=2, mark options={solid}]
table {%
2 1
2.1 1
2.2 1
2.3 1
2.4 1
2.5 1
2.6 0.999
2.7 0.998
2.8 0.992
2.9 0.973
3 0.9435
3.1 0.8645
3.2 0.746
3.3 0.612
3.4 0.41475
3.5 0.269
3.6 0.155375
3.7 0.0738571428571429
3.8 0.0284444444444444
3.9 0.0105104166666667
4 0.00308895705521472
4.1 0.000789889415481833
4.2 0.000174216027874564
4.3 3.37018064168239e-05
4.4 5.32248964775764e-06
4.5 8.00112655861945e-07
4.6 1.29901810097436e-07
4.7 2.34242726060628e-08
4.8 9.0340311955135e-09
};
\addlegendentry{4,4 OMSQ \cite{jinghu_chen_reduced-complexity_2005}}
\end{axis}

\end{tikzpicture}

    \vspace{-0.1cm}
    \caption{Frame error rates for code with rate $R=0.84$.}
    \label{fig:ber_code3}
\end{figure}

Next, we take a look at the frame error rate performance in Fig.~\ref{fig:ber_code3}. The design $E_b/N_0$ is optimized for best performance at frame error rates equal to $10^{-7}$. 
All quantized decoders use 4 bits for the channel message.
Remarkably, even the 2-bit check-node-aware decoder outperforms the 32-bit belief propagation (BP) decoder. Further, the 4-bit check-node-aware decoder achieves similar performance as the non-uniform computational domain design developed in \cite{lee_memory-efficient_2005, he_mutual_2019}. Yet, the complexity of the proposed decoder is significantly smaller by using the minimum approximation in the check node and uniform quantization in the variable node\cite{mohr_uniform_2022}.
The 2-bit decoder exhibits a slightly higher error floor than the decoders which use more bits for the exchanged messages.

\begin{figure}[t]
        \vspace{-0.4cm}
    
    \pgfkeys{/includestandaloneresized, default, width=0.9\linewidth, ratio=0.85}
    \setlength\figurewidth{\itgWidth}
    \setlength\figureheight{\itgRatio\figurewidth}
    \pgfplotsset{legend style={nodes={scale=0.7, transform shape}}}
\pgfplotsset{ every non boxed x axis/.append style={x axis line style=-},
               every non boxed y axis/.append style={y axis line style=-}}
\begin{tikzpicture}

\definecolor{color0}{rgb}{0,0.270588235294118,0.541176470588235}
\definecolor{color1}{rgb}{0.584313725490196,0.729411764705882,0.309803921568627}

\begin{axis}[
height=\figureheight,
legend cell align={left},
legend style={fill opacity=0.8, inner sep=1.5pt, draw opacity=1, text opacity=1, draw=white!80!black, at={(0.99,0.99)}},
log basis y={10},
tick align=outside,
tick pos=left,
width=\figurewidth,
x grid style={white!69.0196078431373!black},
xlabel={\(\displaystyle E_b/N_0\) in dB},
xmajorgrids,
xmin=2, xmax=3.13,
xminorgrids,
xtick style={color=black},
y grid style={white!69.0196078431373!black},
ylabel={Frame Error Rate},
ymajorgrids,
ymin=1e-06, ymax=1,
yminorgrids,
ymode=log,
ytick style={color=black}
]
\addplot [semithick, black, mark=*, mark size=2, mark options={solid}]
table {%
2 0.0129358974358974
2.1 0.00157728706624606
2.2 0.000133049494411921
2.3 6.66431402356263e-06
2.4 1.87793427230047e-07
};
\addlegendentry{BP}
\addplot [semithick, color0, mark=triangle*, mark size=2.5, mark options={solid}]
table {%
2 0.9255
2.1 0.742
2.2 0.42625
2.3 0.163125
2.4 0.0358571428571429
2.5 0.00373880597014925
2.6 0.00023890214797136
2.7 6.97710115401253e-06
2.8 1.00939545287536e-07
2.9 0
};
\addlegendentry{3,2}
\addplot [semithick, color0, dashed, mark=triangle*, mark size=2.5, mark options={solid}, forget plot]
table {%
2 0.9985
2.1 0.9865
2.2 0.924
2.3 0.749
2.4 0.43825
2.5 0.14425
2.6 0.0307352941176471
2.7 0.00357857142857143
2.8 0.000202675314146737
2.9 6.2248448052391e-06
3 8.5815084384833e-08
3.1 0
};
\addplot [semithick, color0, mark=square*, mark size=1.75, mark options={solid}]
table {%
2 0.846
2.1 0.5785
2.2 0.25975
2.3 0.067625
2.4 0.00968269230769231
2.5 0.000754518072289157
2.6 2.96665480004747e-05
2.7 6.30433677360313e-07
2.8 0
};
\addlegendentry{4,2}
\addplot [semithick, color0, dashed, mark=square*, mark size=1.75, mark options={solid}, forget plot]
table {%
2 0.9715
2.1 0.886
2.2 0.635
2.3 0.3355
2.4 0.1062
2.5 0.0176896551724138
2.6 0.00155763239875389
2.7 7.03927917781219e-05
2.8 1.58737717669095e-06
2.9 1.21799712552678e-08
3 0
};
\addplot [semithick, color1, mark=triangle*, mark size=2.5, mark options={solid}]
table {%
2 0.35875
2.1 0.1232
2.2 0.0243333333333333
2.3 0.00256923076923077
2.4 0.000143061516452074
2.5 4.1134126634347e-06
2.6 7.82989004597264e-08
2.7 0
};
\addlegendentry{3,3}
\addplot [semithick, color1, dashed, mark=triangle*, mark size=2.5, mark options={solid}, forget plot]
table {%
2 0.5005
2.1 0.225666666666667
2.2 0.0565555555555556
2.3 0.00776923076923077
2.4 0.000620198265179678
2.5 2.13559931660822e-05
2.6 3.61688568918896e-07
2.7 1.71156676821963e-08
};
\addplot [semithick, color1, mark=square*, mark size=1.75, mark options={solid}]
table {%
2 0.1895
2.1 0.043125
2.2 0.0056875
2.3 0.000369549150036955
2.4 1.29125088761819e-05
2.5 2.31677685483415e-07
2.6 0
};
\addlegendentry{4,3}
\addplot [semithick, color1, dashed, mark=square*, mark size=1.75, mark options={solid}, forget plot]
table {%
2 0.308
2.1 0.0951666666666667
2.2 0.0168225806451613
2.3 0.00140449438202247
2.4 6.45827951433738e-05
2.5 1.27529491194839e-06
2.6 2.34016661986333e-08
};
\addplot [semithick, red!75.2941176470588!black, mark=square*, mark size=1.75, mark options={solid}]
table {%
2 0.0859166666666667
2.1 0.0141388888888889
2.2 0.00122493887530562
2.3 5.97157530156455e-05
2.4 1.4296947601687e-06
2.5 1.61500446817903e-08
2.6 5.10021930943031e-09
};
\addlegendentry{4,4}
\addplot [semithick, red!75.2941176470588!black, dashed, mark=square*, mark size=1.75, mark options={solid}, forget plot]
table {%
2 0.1069
2.1 0.02018
2.2 0.00186567164179104
2.3 0.00010183299389002
2.4 2.46299352232704e-06
2.5 2.63824398480371e-08
2.6 0
};
\addplot [semithick, black, mark options={solid}]
table {%
    0 0
    0 1
};
\addlegendentry{aware}
\addplot [semithick, black, mark options={solid}, dashed]
table {%
    0 0
    0 1
};
\addlegendentry{unaware}
\addplot [semithick, gray, mark=*, mark size=2, mark options={solid}]
table {%
2 0.262
2.1 0.0753571428571429
2.2 0.012047619047619
2.3 0.00094256120527307
2.4 3.95945517896737e-05
2.5 9.32757432648191e-07
2.6 0
};
\addlegendentry{4,4 OMSQ}

\node (A) at (2.6,0.00023890214797136) {};
\node (B) at (2.8,0.00023890214797136) {};
\draw[thick,<->] (A) -- (B) node[midway,sloped,above,rotate=0] {$0.2$\,dB};
\end{axis}

\end{tikzpicture}

    \vspace{-0.1cm}
    \caption{Frame error rates for code with rate $R=0.5$.}
    \label{fig:fer_code6_20}
\end{figure}
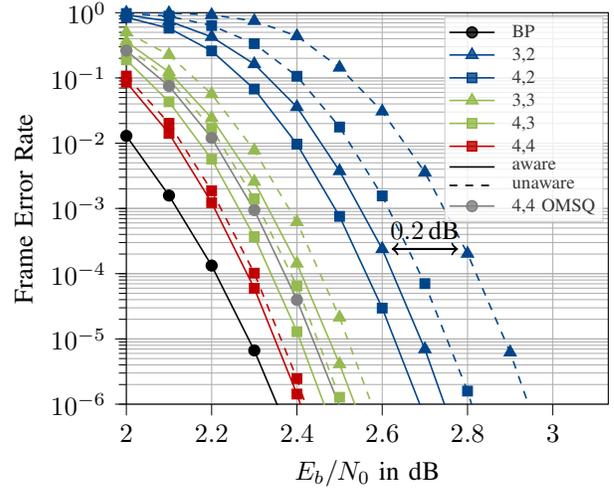

The second example applies a medium-rate regular LDPC code from\cite{mohr_coarsely_2021} with $d_v{=}4$, $d_c{=}8$, $N{=}7648$, $R{=}0.5$ and 20 decoding iterations. Again significant gains of up to 0.2\,dB can be observed in Fig.~\ref{fig:fer_code6_20}, when comparing the check node aware and unaware decoders at same decoding bit width. On the other hand, our results suggest that the performance loss between 2- and 3-bit decoding is larger for medium-rate than for high-rate codes.

\section{Conclusions}
In this paper, we proposed a new class of mutual information maximizing decoders, where the variable node's quantizer design maximizes the mutual information in the check node \emph{output} messages.
The optimization complexity was significantly reduced by restriction to the low-complexity uniform quantization approach in the variable node\cite{mohr_uniform_2022} and the minimum approximation used in the check node\cite{meidlinger_quantized_2015}. 

Simulation results revealed significant decoding performance gains ranging from $0.02$ to 0.2\,dB compared to the conventional optimization, where the variable node's quantizer design maximizes the mutual information in the check node \emph{input} messages\cite{mohr_uniform_2022}.
The highest gains were achieved for very coarse quantization with 2 bits. We also showed that the 2-bit check node operation offered major complexity reduction potential over the 3-bit check node. 
Therefore, we believe that 2-bit decoding of LDPC codes is a promising candidate for applications with low-cost, low-energy and/or high-throughput requirements.

\section{Appendix}
\subsection{Optimization Procedure of the Check Node Aware Design}\label{sec:optimization_procedure}
A grid based search for $\max_{\Delta,r} I(\myvar{X};\myvar{T}^c)$ can be implemented in parallel for a certain range $\Delta\in \mathbb{R}^+$ and a small set of right shifts $r{\in}\mathbb{N}_0$. As a starting point, we assume that the joint distribution of the channel and extrinsic check node messages w.r.t. the code bit $b$ are given by $p(b,t^{ch})$ and $p(b,t^{c}_i),i{\in}\{1,{\ldots},d_v{-}1\}$, respectively. 
After initializing $p(b,\myvar{S}_{0}{=}\phi^{ch}_\Delta(t^{ch})){=} p(b, t^{ch})$, we obtain the output distribution of the summation operation recursively for $k{=}1,\ldots,d_v{-}1$ as
\begin{align}
p(b,s_{k})=\sum_{t^c_{k}} p(b, \myvar{S}_{k-1}{=}s_k-\phi^c_{\Delta,k}(t^{c}_{k}))p(t^{c}_{k}|b).
\end{align}
The uniform quantization $Q(y^v)$, defined in (\ref{equ:uniform_quantization}), performs an $r$-bit right shift operation and a clipping on $y^v{\coloneqq} s_{d_v-1}$ which leads to the output distribution%
\begin{align}
p(b, t^v)=\sum_{y^v}p(t^v|y^v)p(b,y^v).
\end{align}
A symmetry preserving structure like in \cite{mohr_uniform_2022} is assumed where
\begin{align}
p(t^v|y^v)=\begin{dcases}
1/2 & t^v=\pm 1,y^v=0\\
\delta(t^v-Q(y^v)) & otherwise 
\end{dcases}.
\end{align}
At the check node, in case of regular codes, the distributions for extrinsic variable node messages are given by $p(b_i, t^v_i)= p(b, t^v)\forall \, i\in \{1,\ldots, d_{c}{-}1\}$.
The joint distribution computation for the check node message $t^{c}{\coloneqq} t_{d_c-1}$ is initialized with $p(x_{1}, t_{1}){=}p(b_1, t^v_1)$ and performed recursively for $k=2,\ldots,d_c{-}1$ as
\begin{align}
\begin{split}
p(x_{k}, t_{k})&=\sum_{t_{k-1}, t^v_{k}}p(t_k|t_{k-1}, y_{k})p(x_k, t_{k-1}, t^v_{k})
\end{split}
\end{align}
where
\begin{align}
\begin{split}
p(x_k, t_{k-1}, t^v_{k}) 
&=\sum_{b_k}p(\myvar{X}_{k{-}1}{=}x_k{\oplus}b_k,t_{k{-}1}) p(b_k, t^v_k),
\end{split}
\end{align}
\begin{align}
\begin{split}
p(t_k|t_{k-1}, t^v_{k})=&\delta(\operatorname{sgn}(t_k) - \operatorname{sgn}(t_{k-1})\operatorname{sgn}(y_k))\\
&\delta(|t_k| - \min(|t_{k-1}|, |y_k|)).
\end{split}
\end{align}
With $p(x,t^c)=p(x_{d_c}, t_{d_c-1})$ and the Kullback Leibler divergence $D_{\mylabel{KL}}(p(x|t^c)||p(x))=\sum_x p(x|t^c)\log_2(p(x|t^c)/p(x))$, the mutual information yields
\begin{align}
\begin{split}
I(\myvar{X};\myvar{T}^c)%
&=\sum_{t^c} p(t^c) D_{\mylabel{KL}}(p(x|t^c)||p(x)).
\end{split}
\end{align}

\bibliographystyle{MyIEEEtran}
\bibliography{literature}

\begin{thebibliography}{10}
\def\url#1{}
\csname url@samestyle\endcsname
\providecommand{\newblock}{\relax}
\providecommand{\bibinfo}[2]{#2}
\providecommand{\BIBentrySTDinterwordspacing}{\spaceskip=0pt\relax}
\providecommand{\BIBentryALTinterwordstretchfactor}{4}
\providecommand{\BIBentryALTinterwordspacing}{\spaceskip=\fontdimen2\font plus
\BIBentryALTinterwordstretchfactor\fontdimen3\font minus
  \fontdimen4\font\relax}
\providecommand{\BIBforeignlanguage}[2]{{%
\expandafter\ifx\csname l@#1\endcsname\relax
\typeout{** WARNING: IEEEtran.bst: No hyphenation pattern has been}%
\typeout{** loaded for the language `#1'. Using the pattern for}%
\typeout{** the default language instead.}%
\else
\language=\csname l@#1\endcsname
\fi
#2}}
\providecommand{\BIBdecl}{\relax}
\BIBdecl

\bibitem{lewandowsky_trellis_2015}
\BIBentryALTinterwordspacing
J.~Lewandowsky and G.~Bauch, ``\BIBforeignlanguage{en}{Trellis based node
  operations for {LDPC} decoders from the {Information} {Bottleneck} method},''
  in \emph{\BIBforeignlanguage{en}{2015 9th {International} {Conference} on
  {Signal} {Processing} and {Communication} {Systems} ({ICSPCS})}}.\hskip 1em
  plus 0.5em minus 0.4em\relax Cairns, Australia: IEEE, Dec. 2015, pp. 1--10.
  \url{http://ieeexplore.ieee.org/document/7391731/}
\BIBentrySTDinterwordspacing

\bibitem{meidlinger_quantized_2015}
M.~Meidlinger, A.~Balatsoukas-Stimming, A.~Burg, and G.~Matz, ``Quantized
  message passing for {LDPC} codes,'' in \emph{2015 49th {Asilomar} {Conf.} on
  {Signals}, {Systems} and {Computers}}, Nov. 2015, pp. 1606--1610.

\bibitem{he_mutual_2019}
X.~He, K.~Cai, and Z.~Mei, ``On {Mutual} {Information}-{Maximizing} {Quantized}
  {Belief} {Propagation} {Decoding} of {LDPC} {Codes},'' in \emph{2019 {IEEE}
  {Global} {Communications} {Conference} ({GLOBECOM})}, Dec. 2019, pp. 1--6.

\bibitem{mohr_coarsely_2021}
P.~Mohr, G.~Bauch, F.~Yu, and M.~Li, ``Coarsely {Quantized} {Layered}
  {Decoding} {Using} the {Information} {Bottleneck} {Method},'' in \emph{{ICC}
  2021 - {IEEE} {International} {Conference} on {Communications}}, Jun. 2021,
  pp. 1--6.

\bibitem{wang_reconstruction-computation-quantization_2022}
L.~Wang, C.~Terrill, M.~Stark, Z.~Li, S.~Chen, C.~Hulse, C.~Kuo, R.~Wesel,
  G.~Bauch, and R.~Pitchumani, ``Reconstruction-{Computation}-{Quantization}
  ({RCQ}): {A} {Paradigm} for {Low} {Bit} {Width} {LDPC} {Decoding},''
  \emph{IEEE Transactions on Communications}, pp. 1--1, 2022.

\bibitem{mohr_uniform_2022}
P.~Mohr and G.~Bauch, ``Uniform {vs}. {Non}-{Uniform} {Coarse} {Quantization}
  in {Mutual} {Information} {Maximizing} {LDPC} {Decoding},'' in
  \emph{{GLOBECOM} 2022 - 2022 {IEEE} {Global} {Comm.} {Conference}}, Dec.
  2022, pp. 1--6.

\bibitem{jinghu_chen_reduced-complexity_2005}
J.~Chen, A.~Dholakia, E.~Eleftheriou, M.~Fossorier, and X.-Y. Hu,
  ``Reduced-complexity decoding of {LDPC} codes,'' \emph{IEEE Transactions on
  Communications}, vol.~53, no.~8, pp. 1288--1299, Aug. 2005, conference Name:
  IEEE Transactions on Communications.

\bibitem{kurkoski_noise_2008}
\BIBentryALTinterwordspacing
B.~M. Kurkoski, K.~Yamaguchi, and K.~Kobayashi, ``\BIBforeignlanguage{en}{Noise
  {Thresholds} for {Discrete} {LDPC} {Decoding} {Mappings}},'' in
  \emph{\BIBforeignlanguage{en}{{IEEE} {GLOBECOM} 2008 - 2008 {IEEE} {Global}
  {Telecommunications} {Conference}}}.\hskip 1em plus 0.5em minus 0.4em\relax
  New Orleans, LA, USA: IEEE, 2008, pp. 1--5.
  \url{http://ieeexplore.ieee.org/document/4697989/}
\BIBentrySTDinterwordspacing

\bibitem{kschischang_factor_2001}
F.~R. Kschischang, B.~J. Frey, and H.~Loeliger, ``Factor graphs and the
  sum-product algorithm,'' \emph{IEEE Transactions on Information Theory},
  vol.~47, no.~2, pp. 498--519, Feb. 2001.

\bibitem{xiao-yu_hu_efficient_2001}
\BIBentryALTinterwordspacing
{Xiao-Yu Hu}, E.~Eleftheriou, D.-M. Arnold, and A.~Dholakia,
  ``\BIBforeignlanguage{en}{Efficient implementations of the sum-product
  algorithm for decoding {LDPC} codes},'' in
  \emph{\BIBforeignlanguage{en}{{GLOBECOM}'01. {IEEE} {Global}
  {Telecommunications} {Conference} ({Cat}. {No}.{01CH37270})}}, vol.~2.\hskip
  1em plus 0.5em minus 0.4em\relax San Antonio, TX, USA: IEEE, 2001, pp.
  1036--1036E.  \url{http://ieeexplore.ieee.org/document/965575/}
\BIBentrySTDinterwordspacing

\bibitem{lee_low-complexity_2015}
\BIBentryALTinterwordspacing
Y.~Lee, B.~Kim, J.~Jung, and I.-C. Park,
  ``\BIBforeignlanguage{en}{Low-{Complexity} {Tree} {Architecture} for
  {Finding} the {First} {Two} {Minima}},'' \emph{\BIBforeignlanguage{en}{IEEE
  Trans. on Circ. and Systems II: Express Briefs}}, vol.~62, no.~1, pp. 61--64,
  Jan. 2015.  \url{http://ieeexplore.ieee.org/document/6922514/}
\BIBentrySTDinterwordspacing

\bibitem{kurkoski_quantization_2014}
B.~M. Kurkoski and H.~Yagi, ``Quantization of {Binary}-{Input} {Discrete}
  {Memoryless} {Channels},'' \emph{IEEE Transactions on Information Theory},
  vol.~60, no.~8, pp. 4544--4552, Aug. 2014.

\bibitem{noauthor_standard_2006}
``Standard for {Information} {Technology} - {Telecommunications} and
  {Information} {Exchange} {Between} {Systems} – {LAN}/{MAN} - {Specific}
  {Requirements} {Part} 3: {CSMA}/{CD} {Access} {Method} and {Physical} {Layer}
  {Specifications} - {Amendment}: {Physical} {Layer} and {Management}
  {Parameters} for 10 {Gb}/s {Operation}, {Type} {10GBASE}-{T},'' \emph{IEEE
  Std 802.3an-2006 (Amendment to IEEE Std 802.3-2005)}, pp. 1--181, Sep. 2006,
  conference Name: IEEE Std 802.3an-2006 (Amendment to IEEE Std 802.3-2005).

\bibitem{lee_memory-efficient_2005}
\BIBentryALTinterwordspacing
J.-S. Lee and J.~Thorpe, ``\BIBforeignlanguage{en}{Memory-efficient decoding of
  {LDPC} codes},'' in \emph{\BIBforeignlanguage{en}{Proceedings.
  {International} {Symposium} on {Information} {Theory}, 2005. {ISIT}
  2005.}}\hskip 1em plus 0.5em minus 0.4em\relax Adelaide, Australia: IEEE,
  2005, pp. 459--463.  \url{http://ieeexplore.ieee.org/document/1523376/}
\BIBentrySTDinterwordspacing

\end{thebibliography}
\newpage
\newpage

\end{document}